\documentclass[twocolumn]{IEEEtran}
\usepackage{color}
\usepackage{paralist}
\usepackage{booktabs, multirow}
\usepackage{pstool}
\usepackage{amsmath,amsfonts,amssymb}
\usepackage{mathabx,fixmath}
\usepackage{algorithm}
\usepackage{algorithmic}
\usepackage{cite}
\usepackage[lowtilde]{url}
\def\x{\mathbold{x}}
\def\A{\mathbold{A}}

\def\r{\mathbold{r}}

\def\y{\mathbold{y}}

\renewcommand{\b}{\mathbold{b}}

\def\qed{\hfill $\blacksquare$ }

\newtheorem{remark}{Remark}

\newtheorem{proposition}{Proposition}
\newtheorem{assumption}{Assumption}

\newtheorem{theorem}{Theorem}
\newtheorem{corollary}{Corollary}
\input{mysymbol.sty}

\title{A Class of Prediction-Correction Methods for Time-Varying Convex Optimization}

\author{Andrea~Simonetto$^*$,~Aryan~Mokhtari$^\dagger$,~Alec~Koppel$^\dagger$,~Geert~Leus$^*$, and~Alejandro~Ribeiro$^\dagger$

\thanks{The work in this paper is supported in part by STW under the D2S2 project from the ASSYS program (project 10561) and in part by NSF CAREER CCF-0952867, and ONR N00014-12-1-0997. This paper expands the results and presents convergence proofs that are referenced in~\cite{Paper1Asilomar} and \cite{Paper2GlobalSip}. }
\thanks{
$^*$Andrea Simonetto and Geert Leus are with the Department of Electrical Engineering, Mathematics and Computer Science, Delft University of Technology, 2826 CD Delft, The Netherlands. Email: \{a.simonetto, g.j.t.leus\}@tudelft.nl. }
\thanks{
$^\dagger$Aryan Mokhtari, Alec Koppel, and Alejandro Ribeiro are with the Department of Electrical and Systems Engineering, University of Pennsylvania, 200 South 33rd Street, Philadelphia, PA 19104, USA. Email: \{aryanm, akoppel, aribeiro\}@seas.upenn.edu. }
}

\begin{document}

\maketitle

\begin{abstract}%
This paper considers unconstrained convex optimization problems with time-varying objective functions. We propose algorithms with a discrete time-sampling scheme to find and track the solution trajectory based on prediction and correction steps, while sampling the problem data at a constant rate of $1/h$, where $h$ is the sampling period. The prediction step is derived by analyzing the iso-residual dynamics of the optimality conditions. The correction step adjusts for the distance between the current prediction and the optimizer at each time step, and consists either of one or multiple gradient steps or Newton steps, which respectively correspond to the gradient trajectory tracking (GTT) or Newton trajectory tracking (NTT) algorithms. Under suitable conditions, we establish that the asymptotic error incurred by both proposed methods behaves as $O(h^2)$, and in some cases as $O(h^4)$, which outperforms the state-of-the-art error bound of $O(h)$ for correction-only methods in the gradient-correction step. Moreover, when the characteristics of the objective function variation are not available, we propose approximate gradient and Newton tracking algorithms (AGT and ANT, respectively) that still attain these asymptotical error bounds. Numerical simulations demonstrate the practical utility of the proposed methods and that they improve upon existing techniques by several orders of magnitude. 
\end{abstract}%



\section{Introduction}

In this paper, we consider unconstrained optimization problems whose objective functions vary continuously in time. In particular, consider a variable $\x\in \reals^n$ and a non-negative continuous time variable $t\in \reals_+$, which determine the choice of a \emph{smooth strongly convex function} $f: \mathbb{R}^n\times \mathbb{R}_{+} \to \mathbb{R}$. We study the problem
\begin{equation}\label{eq.problem}
\x^*(t) :=\argmin_{\x\in \mathbb{R}^n} f(\x; t), \quad \textrm{for } t \geq 0 \;.
\end{equation}
%
Our goal is to determine the solution $\x^*(t)$ of~\eqref{eq.problem} for each time $t$ which corresponds to the solution \emph{trajectory}. Time-varying optimization problems of the form \eqref{eq.problem} arise in control  \cite{Hours2014,castillo2007unmanned,vahidi2003research}, when, for instance, one is interested in generating a control action such that the system remains close to a dynamical reference trajectory, as well as in signal processing \cite{Jakubiec2013}, where one seeks to estimate a dynamical process based on time-varying observations. Other examples arise in robotics\cite{ardeshiri2010convex,verscheure2009time,KoppelEtal15a,cKoppelEtal15a,cKoppelEtal15b} and economics~\cite{Dontchev2013}.

The problem in \eqref{eq.problem} can be solved based on a continuous time
platform \cite{zhao1993training,myung1997time,baumann2004newton,fazlyab2015interior} or can be interpreted as a sequence of time-invariant problems. In particular, one could sample the objective functions $ f(\x; t)$ at time instants $t_{k}$ with $k=0,1,2,\dots$, and sampling period $h = t_k - t_{k-1}$, arbitrarily close to each other and then solve the resulting time-invariant problems
\begin{equation}\label{eq.problemd}
\x^*(t_k) :=\argmin_{\x\in \mathbb{R}^n} f(\x; t_k).
\end{equation}
By decreasing $h$, an arbitrary accuracy may be achieved when approximating \eqref{eq.problem} by \eqref{eq.problemd}. However, solving \eqref{eq.problemd} for each sampling time $t_k$ is not a viable option in most application domains, even for moderate-size problems. The requisite computation time for solving each instance of the problem often does not meet the requirements for real-time applicability, as in the control domain \cite{Bittanti2001}. 
%
%
It is also challenging to reasonably bound the time each problem instance will take to be solved~\cite{Boyd2004a}. In short, the majority of iterative methods for convex problems with static objectives may not be easily extended to handle time-varying objectives, with the exception of when the changes in the objective occur more slowly than the time necessary for computing the optimizer. 

Instead, we consider using the tools of \emph{non-stationary} optimization \cite{Gupal1974, Kheisin1976, Nurminskii1977, Ermoliev1982}\cite[Chapter~6]{Polyak1987} to solve problems of the form \eqref{eq.problem}. In these works the authors consider perturbations of the time-varying problem when an initial solution $\x^*(t_0)$ is known. More recently, the work presented in~\cite{Popkov2005} designs a gradient method for unconstrained optimization problems using an arbitrary starting point, which achieves a $\|\x(t_k) - \x^*(t_k) \|=O(h)$ asymptotic error bound with respect to the optimal trajectory.
Time-varying optimization has also been studied in the context of \emph{parametric programming}, where the optimization problem is parametrized over a parameter vector $\mathbold{p}\in\mathbb{R}^p$ that may represent time, as studied in~\cite{Robinson1980,Dontchev2009,Guddat1990}. Tracking algorithms for optimization problems with parameters that change in time are given in~\cite{Zavala2010,Dontchev2013} and are based on predictor-corrector schemes. 
Even though these algorithms are applicable to constrained problems, they assume the access to an initial solution $\x^*(t_0)$, which may not be available in practice. Some of the theoretical advances in these works have been used to ease the computational burden of sequential convex programming while solving nonconvex optimization problems, or nonlinear model predictive control~\cite{Hours2014,Dinh2012,Diehl2005}. 

In this paper, we design iterative discrete-time sampling algorithms initialized at an arbitrary point $\x_0$ which converge asymptotically to the solution trajectory $\x^*(t)$ up to an error bound which may be specified as arbitrarily small and depends on the sampling period $h$. In particular, the methods proposed here yield a sequence of approximate time-varying optimizers $\{\x_k\}$, for which
%
$\limsup_{k\to \infty} \|\x_k -\x^*(t_k)\| \leq \delta$
%
with $\delta$ dependent on the sampling period $h$. To do so, we predict where the optimal continuous-time trajectory will be at the next sampling time and track the associated prediction error based upon estimating the curvature of the solution trajectory. Under suitable assumptions, we establish that the proposed prediction-correction scheme attains an asymptotic error bound of $O(h^2)$ and in some cases $O(h^4)$, which outperforms the $O(h)$ error bound achieved by the state-of-the-art method of~\cite{Popkov2005}.%

In Section \ref{sec:probform}, we analyze unconstrained optimization problems and we propose algorithms to track their time-varying solution which fall into the family of tracking algorithms with \emph{an arbitrary} starting point.The proposed methods are based on a predictor-corrector approach, where the predictor step is generated via a Taylor expansion of the optimality conditions, and the correction step may either be a single or multiple gradient descent or Newton steps. In Section \ref{sec:conv}, we show that our tracking methods converge to the solution trajectory asymptotically, with an error bound $O(h^2)$ (and in some cases $O(h^4)$ locally) dependent on the sampling period $h$. This error bound improves upon the existing methods which attain an $O(h)$ bound. We further extend the tracking framework to account for the case where the dependence of the cost function on the time parameter is not known a priori but has to be estimated, and establish that the $O(h^2)$ and the (local) $O(h^4)$ asymptotical error bound are achieved despite the associated estimation uncertainty. In Section \ref{sec:num} we numerically analyze the performance of the proposed methods as compared with existing approaches. In particular, in Section \ref{sec:sims} we consider a scalar example and show the convergence bounds hold in practice, and in Section \ref{sec:robots} we apply the proposed method to a reference path following problem and use the tools developed here to yield an effective control strategy for an intelligent system. Finally, in Section \ref{sec:conclusions} we close the paper by concluding remarks.

{\bf Notation.} Vectors are written as $\x\in\reals^n$ and matrices as $\A\in\reals^{n\times n}$. We use $\|\cdot\|$ to denote the Euclidean norm, both in the case of vectors, matrices, and tensors. The gradient of the function $f(\x; t)$ with respect to $\x$ at the point $(\x,t)$ is indicated as $\nabla_{\x} f(\x; t) \in \reals^n$, while the partial derivative of the same function w.r.t. $t$ at $(\x,t)$ is written as $\nabla_t f(\x; t)\in \reals$. Similarly, the notation $\nabla_{\x\x} f(\x; t) \in \reals^{n\times n}$ denotes the Hessian of $f(\x;t)$ w.r.t. $\x$ at $(\x,t)$, whereas $\nabla_{t\x} f(\x; t) \in \reals^{n}$ denotes the partial derivative of the gradient of $f(\x;t)$ w.r.t. the time $t$ at $(\x,t)$, i.e. the mixed first-order partial derivative vector of the objective. The tensor $\nabla_{\x\x\x} f(\x; t) \in \reals^{n\times n\times n}$ indicates the third derivative of $f(\x;t)$ w.r.t. $\x$ at $(\x,t)$, the matrix $\nabla_{\x t\x} f(\x; t) = \nabla_{t\x\x} f(\x; t)   \in \reals^{n\times n}$ indicates the time derivative of the Hessian of $f(\x;t)$ w.r.t. the time $t$ at $(\x,t)$, and the vector $\nabla_{t t\x} f(\x; t) \in \reals^{n}$ indicates the second derivative in time of the gradient of $f(\x;t)$ w.r.t. the time $t$ at $(\x,t)$.


\section{Algorithm definition }\label{sec:probform}

In this section we introduce a class of algorithms for solving optimization problem \eqref{eq.problem} using prediction and correction steps. In order to converge to the solution trajectory $\x^*(t)$, we generate a sequence of near optimal decision variables $\{\x_k\}$ by taking into account both how the solution changes in time and how different our current update is from the optimizer at each time step. 

\subsection{Gradient trajectory tracking}
In this paper we assume that the initial decision variable $\x_0$ is not necessarily the optimal solution of the initial objective function $f(\x;t_0)$, i.e., $\x_0\neq \x^*(t_0)$. We model this assumption by defining a residual error for the gradient of the initial variable $\nabla_\x f(\x_0;t_0)=\r(0)$. To improve the estimation for the decision variable $\x$, we set up a prediction-correction scheme motivated by the Kalman filter strategy in estimation theory~\cite{Bar-Shalom2001} and by continuation methods in numerical analysis~\cite{Allgower1990}. In the first step, we predict how the solution changes, and in the correction step we use descent methods to push the predicted variable towards the optimizer at that time instance\footnote{This correction strategy has been called differently by different authors: an alternative term is \emph{adaptation}, as reported in~\cite{Sayed2014,Zhao2015}.}.

To generate the prediction step, we reformulate the time-varying problem~\eqref{eq.problem} in terms of its optimality conditions. Minimizing the objective in~\eqref{eq.problem} is equivalent to computing the solution of the following nonlinear system of equations
\begin{equation}\label{equivalent_problem}
\nabla_{\x} f(\x^*(t); t) = \mathbf{0}, 
\end{equation}
for each $t$. These two problems are equivalent since the objective functions $f(\x; t) $ are strongly convex with respect to $\x$ and only their optimal solutions satisfy the condition in~ \eqref{equivalent_problem}. 

Consider an arbitrary vector $\x\in \reals^n$ which may be interpreted as the state of a dynamical system. The objective function gradient $\nabla_\x f({\x};t)\in \reals^n$ computed at point ${\x}$ is
\begin{equation}\label{eq.isoerror}
\nabla_{\x} f({\x}; t) = \r(t),
\end{equation}
where $\r(t)\in \reals^n$ is the residual error. 
The aim of the prediction step is to keep the residual error as constant as possible while the optimization problem is changing. To say it in another way, we want to predict how to update $\x_k$ such that we stay close to the iso-residual manifold.  We try to keep the evolution of the trajectory close to the residual vector $\r(t)$ which is equivalent to
\begin{multline}\label{variation_formula}
\nabla_{\x} f (\x + \delta \x; t + \delta t) \approx \\
\nabla_{\x} f(\x; t) + \nabla_{\x\x} f(\x; t) \delta\x + \nabla_{t\x} f(\x; t) \delta t = \r(t),  
\end{multline}
where ${{\delta \x}}\in\reals^n$ and the positive scalar ${{\delta t}}$ are the variations of the decision variable $\x$ and the time variable $t$, respectively. By subtracting~\eqref{eq.isoerror} from~\eqref{variation_formula} and dividing the resulting equation by the time variation $\delta t$, we obtain the continuous dynamical system
\begin{equation}\label{eq.dyn}
\dot{\x} = - [\nabla_{\x\x} f(\x; t)]^{-1} \nabla_{t\x} f(\x; t),
\end{equation}
where ${{\dot{\x}:= \delta \x/\delta t}}$. We then consider the discrete time approximation of \eqref{eq.dyn}, which amounts to sampling the problem at times $t_k$, for $k = 0,1,2,\dots$\ . The prediction step consists of a discrete-time approximation of integrating~\eqref{eq.dyn} by using an Euler scheme. Let ${{\x_{k+1|k}\in \reals^n}}$ be the predicted decision variable based on the available information up to time $t_k$, then we may write the Euler integral approximation of \eqref{eq.dyn} as
\begin{equation}\label{GTT_prediction}
{\x}_{k+1|k} = \x_k - h\,[\nabla_{\x\x} f(\x_k; t_k)]^{-1} \nabla_{t\x} f(\x_k; t_k).
\end{equation}
Observe that the prediction step in \eqref{GTT_prediction} is computed by only incorporating information available at time $t_k$; however, the decision variable ${\x}_{k+1|k}$ is supposed to be close to the iso-residual manifold of the objective function at time $t_{k+1}$. 

The gradient trajectory tracking (GTT) algorithm uses the  gradient descent method to correct the predicted decision variable ${\x}_{k+1|k}$. This procedure modifies the predicted variable ${\x}_{k+1|k}$ towards the optimal argument of the objective function at time $t_{k+1}$. Therefore, the correction (or adaptation) step of GTT requires execution of the gradient descent method based on the updated objective function $f(\x;t_{k+1})$. Depending on the sampling period $h$, we can afford a specific number of gradient descent steps until sampling the next function. 

Define $\tau$ as the number of gradient descent steps used for correcting the predicted decision variable ${\x}_{k+1|k}$. Further, define $\hat{\x}_{k+1}^s \in \reals^n$ as the corrected decision variable after executing $s$ steps of the gradient descent method. Therefore, the sequence of variables $\hat{\x}_{k+1}^s$ is initialized by $\hat{\x}_{k+1}^0={\x}_{k+1|k}$ and updated by the recursion
\begin{equation}\label{GTT_correction}
\hat{\x}_{k+1}^{s+1}= \hat{\x}_{k+1}^s- \gamma \nabla_{\x} f(\hat{\x}_{k+1}^s; t_{k+1}),
\end{equation}
where $\gamma > 0$ is the stepsize. The output of the recursive update \eqref{GTT_correction} after $\tau$ steps is the decision variable of the GTT algorithm at time $t_{k+1}$, i.e., $\x(t_{k+1}):=\x_{k+1}=\hat{\x}_{k+1}^{\tau}$.


{\begin{algorithm}[tb]
\caption{Gradient trajectory tracking (GTT)}\label{algo_GTT} 
\begin{algorithmic}[1] 
\small{\REQUIRE  Initial variable $\x_0$. Initial objective function $f(\x;t_0)$, no. of correction steps $\tau$
\FOR {$k=0,1,2,\ldots$}
   \STATE Predict the solution using the prior information [cf \eqref{GTT_prediction}]  
  $$\displaystyle{
{\x}_{k+1|k} = \x_k - h\, [\nabla_{\x\x} f(\x_k; t_k)]^{-1} \nabla_{t\x} f(\x_k; t_k) \, 
}$$
\STATE Acquire the updated function $f(\x;t_{k+1})$
   \STATE Initialize the sequence of corrected variables $ \hat{\x}_{k+1}^0={\x}_{k+1|k}$
          \FOR {$s=0:\tau-1$}
          \STATE  Correct the variable by the  gradient step  [cf \eqref{GTT_correction}]   
         $$\displaystyle{   \quad
\hat{\x}_{k+1}^{s+1}= \hat{\x}_{k+1}^s- \gamma \nabla_{\x} f(\hat{\x}_{k+1}^s; t_{k+1})}$$
          \ENDFOR
   \STATE Set the corrected variable $\displaystyle{{\x}_{k+1}=\hat{\x}_{k+1}^{\tau}}$
\ENDFOR}
\end{algorithmic}\end{algorithm}
%


We summarize the GTT scheme in Algorithm \ref{algo_GTT}. Observe that Step 2 and Step 6 implement the prediction-correction scheme. In Step 2, we compute a first-order approximation of the gradient $\nabla_{\x} f(\x; t)$ at time $t_k$ [cf. \eqref{GTT_prediction}]. Then we correct the predicted solution by executing $\tau$ gradient descent steps as stated in \eqref{GTT_correction} for the updated objective function $ f(\x; t_{k+1})$ in Steps 5-7. The sequence of corrected variables is initialized by the predicted solution $\hat{\x}_{k+1}^0={\x}_{k+1|k}$ in Step 4 and the output of the recursion is considered as the updated variable $\x_{k+1}=\hat{\x}_{k+1}^{\tau}$ in Step 8. The implementation of gradient descent for the correction process requires access to the updated function $ f(\x; t_{k+1})$ which is sampled in Step 3.

Note that the GTT correction step is done by executing $\tau$ gradient descent  steps which only uses first-order information of the objective function $f$. We accelerate this procedure using second-order information in the following subsection.


\subsection{Newton trajectory tracking}\label{sec:newton}

The GTT prediction step introduced in \eqref{GTT_prediction} requires computation of the partial Hessian inverse $ [\nabla_{\x\x} f(\x_k; t_k)]^{-1}$. 
Note that the computational complexity of the Hessian inverse is of order $O(n^3)$, which is affordable when $n$ is of moderate size or a certain level of latency associated with this inverse computation will not degrade performance. These two observations justify using the Newton method for the correction (or adaptation) step as well, which requires computation of the partial Hessian inverse of the objective function. Therefore, we introduce the Newton trajectory tracking (NTT) method as an algorithm that uses second-order information for both the prediction and correction steps.  

The prediction step of the NTT algorithm is identical to the prediction step of the GTT method as introduced in \eqref{GTT_prediction}; however, in the correction steps NTT updates the predicted solution trajectory by applying $\tau$ steps of the Newton method. In particular, the predicted variable ${\x}_{k+1|k}$ in \eqref{GTT_prediction} is used for initializing the sequence of corrected variables $\hat{\x}_{k+1}^s$, i.e., $\hat{\x}_{k+1}^0:={\x}_{k+1|k}$. The sequence of corrected variables $\hat{\x}_{k+1}^s$ is updated using Newton steps as
\begin{equation}\label{NTT_correction}
\hat{\x}_{k+1}^{s+1} \!=\! \hat{\x}_{k+1}^s\!- \! \nabla_{\x\x} f(\hat{\x}_{k+1}^s; t_{k+1})^{-1}  \nabla_{\x} f(\hat{\x}_{k+1}^s;  t_{k+1}).
\end{equation}
The decision variable (solution) at step $t_{k+1}$ for the NTT algorithm $\x(t_{k+1}):=\x_{k+1}$ is the outcome of $\tau$ iterations of \eqref{NTT_correction} such that $\x_{k+1}=\hat{\x}_{k+1}^{\tau}$.

Observe that the computational time of the Newton step and the gradient descent step are different. The complexity of the Newton step is in the order of $O(n^3)$, while the gradient descent step requires a computational complexity of order $O(n)$. Since the sampling period is a fixed value, the number of Newton iterations in one iteration of the NTT algorithm is smaller than the number of gradient descent steps that we can afford in the correction step of GTT. On the other hand, the Newton method requires less iterations relative to the gradient descent method to achieve a comparable accuracy. In particular, for an optimization problem with a large condition number the difference between the convergence speeds of these algorithms is substantial, in which case NTT is preferable to GTT.

In developing the prediction steps of the GTT and NTT algorithms we assumed that the mixed partial derivative $\nabla_{t\x} f(\x; t)$ is available; however, frequently in applications the variation of the objective function over time is not known. This motivates the idea of approximating the objective function variation which we study in the following subsection.

{\begin{algorithm}[tb]
\caption{Newton trajectory tracking (NTT)}\label{algo_NTT} 
\begin{algorithmic}[1] 
\small{\REQUIRE  Initial variable $\x_0$. Initial objective function $f(\x;t_0)$, no. of correction steps $\tau$
\FOR {$k=0,1,2,\ldots$}
   \STATE Predict the solution using the prior information [cf \eqref{GTT_prediction}]  
  $$\displaystyle{
{\x}_{k+1|k} = \x_k - h \,[\nabla_{\x\x} f(\x_k; t_k)]^{-1} \nabla_{t\x} f(\x_k; t_k) 
}$$
\STATE Acquire the updated function $f(\x;t_{k+1})$
   \STATE Initialize the sequence of corrected variables $ \hat{\x}_{k+1}^0={\x}_{k+1|k}$
          \FOR {$s=0:\tau-1$}
          \STATE  Correct the variable by the  Newton step  [cf \eqref{GTT_correction}]   
        $$\displaystyle{ 
\hat{\x}_{k+1}^{s+1}= \hat{\x}_{k+1}^s-  \! \nabla_{\x\x} f(\hat{\x}_{k+1}^s; t_{k+1})^{-1} \nabla_{\x} f(\hat{\x}_{k+1}^s; t_{k+1})}$$
          \ENDFOR
   \STATE Set the corrected variable $\displaystyle{{\x}_{k+1}=\hat{\x}_{k+1}^{\tau}}$
\ENDFOR}
\end{algorithmic}\end{algorithm}
%

\subsection{Time derivative approximation}\label{sec:app}

 Consider the mixed partial derivative at time $t_{k}$ using the gradient of the objective with respect to $\x$ at times $t_{k}$ and $t_{k-1}$, that is, the approximate partial mixed gradient $\tilde{\nabla}_{t\x}{f}_k$ as
\begin{equation}\label{fobd}
\tilde{\nabla}_{t\x}f(\x_k; t_k) = \frac{1}{h} \left(\nabla_{\x}f(\x_k; t_k) - \nabla_{\x}f(\x_{k}; t_{k-1})\right). 
\end{equation}
which is called a \textit{first-order backward finite difference} since it requires information of the first previous step for approximating the current mixed partial derivative. The error of this approximation is bounded on the order of $O(h)$~\cite{Quarteroni2000}, which may be improved by using the gradients and mixed partial derivative $\tilde{\nabla}_{t\x}f(\x_k; t_k)$  of more than one previous step, if needed\footnote{Approximation errors of the order of $O(h^2)$, $O(h^3)$, and $O(h^4)$ can be achieved, e.g., by the recursive method presented in~\cite{Lee1992}.}.

Substituting the partial mixed gradient $\nabla_{t\x} f(\x_k; t_k)$ in \eqref{GTT_prediction} by its approximation $\tilde{\nabla}_{t\x}f(\x_k; t_k) $ in \eqref{fobd} leads to the {\it{approximate prediction step}}
\begin{equation}\label{appro_prediction}
{\x}_{k+1|k} = \x_k - h\,[\nabla_{\x\x} f(\x_k; t_k)]^{-1}\tilde{\nabla}_{t\x}f(\x_k; t_k) .
\end{equation}
The predicted variable ${\x}_{k+1|k}$ is an initial estimate for the optimal solution of the objective function $f(\x; t_{k+1})$. This estimation can be corrected by descending through the optimal argument of the objective function $f(\x; t_{k+1})$. To do so, one may either use a gradient algorithm as in \eqref{GTT_correction} or Newton steps as in \eqref{NTT_correction}. Based on this idea, we introduce the approximate gradient tracking (AGT) algorithm which is different from GTT in using the approximate prediction step in \eqref{appro_prediction} instead of the exact update in \eqref{GTT_prediction}. Likewise, we introduce the approximate Newton tracking (ANT) method as a variation of the NTT algorithm. We summarize the AGT and ANT methods which make use of this approximation scheme in Algorithms \ref{algo_AGT} and \ref{algo_ANT}, respectively. As we can observe, the main difference with Algorithms~\ref{algo_GTT} and~\ref{algo_NTT} is in Step 2, where we use the approximate time derivative. In Section \ref{sec:conv} we establish that this time derivative approximation does not degrade significantly the performance of the algorithms presented here. 


{\begin{algorithm}[tb]
\caption{Approximate gradient tracking (AGT)}\label{algo_AGT} 
\begin{algorithmic}[1] 
\small{\REQUIRE  Initial variable $\x_0$. Initial objective function $f(\x;t_0)$, no. of correction steps $\tau$
\FOR {$k=0,1,2,\ldots$}
   \STATE Predict the solution using the prior information [cf. \eqref{GTT_prediction}-\eqref{fobd}]  
  $$\displaystyle{
{\x}_{k+1|k} = \x_k - [\nabla_{\x\x} f(\x_k; t_k)]^{-1} \tilde{\nabla}_{t\x} f(\x_k; t_k) \, h
}$$
\STATE Acquire the updated function $f(\x;t_{k+1})$
   \STATE Initialize the sequence of corrected variables $ \hat{\x}_{k+1}^0={\x}_{k+1|k}$
          \FOR {$s=0:\tau-1$}
          \STATE  Correct the variable by the  gradient step  [cf. \eqref{GTT_correction}]   
         $$\displaystyle{   \quad
\hat{\x}_{k+1}^{s+1}= \hat{\x}_{k+1}^s- \gamma \nabla_{\x} f(\hat{\x}_{k+1}^s; t_{k+1})}$$
          \ENDFOR
   \STATE Set the corrected variable $\displaystyle{{\x}_{k+1}=\hat{\x}_{k+1}^{\tau}}$
\ENDFOR}
\end{algorithmic}\end{algorithm}
%



\section{Convergence Analysis}\label{sec:conv}

We turn to establishing that the prediction-correction schemes derived in Section \ref{sec:probform} solve the continuous-time problem stated in \eqref{eq.problem} up to an error term which is dependent on the discrete-time sampling period. In order to do so, some technical conditions are required which we state below.


\begin{assumption} \label{as.str} The function $f(\x; t)$ is twice differentiable and $m$-strongly convex in $\x\in \mathbb{R}^n$ and uniformly in $t$, that is, the Hessian of $f(\x; t)$ with respect to $\x$ is bounded below by $m$ for each $\x\in \mathbb{R}^n$ and uniformly in $t$, 
$$
\nabla_{\x\x} f(\x; t) \succeq m \mathbf{I}, \quad \forall \x\in \mathbb{R}^n, t.
$$  
\end{assumption}
%

{\begin{algorithm}[tb]
\caption{Approximate Newton tracking (ANT)}\label{algo_ANT} 
\begin{algorithmic}[1] 
\small{\REQUIRE  Initial variable $\x_0$. Initial objective function $f(\x;t_0)$, no. of correction steps $\tau$
\FOR {$k=0,1,2,\ldots$}
   \STATE Predict the solution using the prior information [cf. \eqref{GTT_prediction}-\eqref{fobd}]  
  $$\displaystyle{
{\x}_{k+1|k} = \x_k - [\nabla_{\x\x} f(\x_k; t_k)]^{-1} \tilde{\nabla}_{t\x} f(\x_k; t_k) \, h
}$$
\STATE Acquire the updated function $f(\x;t_{k+1})$
   \STATE Initialize the sequence of corrected variables $ \hat{\x}_{k+1}^0={\x}_{k+1|k}$
          \FOR {$s=0:\tau-1$}
          \STATE  Correct the variable by the Newton step  [cf. \eqref{GTT_correction}]   
         \begin{equation*}
\hat{\x}_{k+1}^{s+1}= \hat{\x}_{k+1}^s- \ \!  \nabla_{\x\x} f(\hat{\x}_{k+1}^s; t_{k+1})^{-1}  \nabla_{\x} f(\hat{\x}_{k+1}^s; t_{k+1})\end{equation*}
          \ENDFOR
   \STATE Set the corrected variable $\displaystyle{{\x}_{k+1}=\hat{\x}_{k+1}^{\tau}}$
\ENDFOR}
\end{algorithmic}\end{algorithm}

\begin{assumption} \label{as.smooth} The function $f(\x; t)$ is sufficiently smooth both in $\x\in \mathbb{R}^n$ and in $t$, and in particular, $f(\x; t)$ has bounded second and third order derivatives with respect to $\x\in \mathbb{R}^n$ and $t$ as
$$
\|\nabla_{\x\x} f(\x; t)\|\leq L, \, \|\nabla_{t\x} f(\x; t)\|\leq C_0, \, \|\nabla_{\x\x\x} f(x; t)\|\leq C_1,
$$
$$
\|\nabla_{\x t\x} f(\x; t)\|\leq C_2, \quad \|\nabla_{t t\x} f(\x; t)\|\leq C_3.
$$ 
\end{assumption}

%
Assumption~\ref{as.str}, besides guaranteeing that problem~\eqref{eq.problem} is strongly convex and has a \emph{unique} solution for each time instance, is needed to ensure that the Hessian of the objective function $f(\x; t)$ is invertible. The fact that the solution is unique for each time instance, implies that the solution trajectory is unique. This mathematical setting frequently appears in the analysis of optimization tools in time-varying settings, and is essential to establishing trajectory tracking results-- see, for instance~\cite{Popkov2005, Dontchev2013, Jakubiec2013, Ling2013}. Assumption~\ref{as.smooth} ensures that the Hessian is bounded from above, a property which is equivalent to the Lipschitz continuity of the gradient, and that the third derivative tensor $\nabla_{\x\x\x} f(x; t)$ is also bounded above (typically required for the analysis of Newton-type algorithms), as well as boundedness of the time variations of gradient and Hessian. These last properties ensure the possibility to build a prediction scheme based on the (estimated) knowledge of how the function and its derivatives change in time. A similar assumption was required (albeit only locally) for the local convergence analysis in~\cite[Eq.~(3.2)]{Dontchev2013}.

Assumptions~\ref{as.str} and ~\ref{as.smooth} are sufficient to show that the solution \emph{mapping} $t \mapsto \x^*(t)$ is single-valued and locally Lipschitz continuous in $t$, and in particular, 
\begin{equation}\label{eq.lip}
\|\x^*(t_{k+1}) \!-\! \x^*(t_{k})\| \leq \! \frac{1}{m}\|\nabla_{t\x} f(\x; t)\| (t_{k+1}\!-\!t_{k}) \leq \frac{C_0 h }{m}, 
\end{equation} 
see for example~\cite[Theorem 2F.10]{Dontchev2009}. This gives us a link between the sampling period $h$ and the allowed variations in the optimizers.  This property also allows our algorithms to converge to a neighborhood of the optimal solution. We remark that, in most of the current literature the condition in~\eqref{eq.lip} is taken as an assumption (that is, one assumes that the optimizer does not change more than a certain upper bound in time), while here is a consequence of our smoothness and boundedness assumptions.

\begin{remark}\label{remark.as}
Assumptions~\ref{as.str} and \ref{as.smooth} can be weakened if a priori knowledge of the domain of the the optimizers and the sequence generated by the algorithms is given by the structure of the problem, i.e. the optimal trajectory is contained within a subset $X$ of $\mathbb{R}^n$.  In this case, we can concentrate on functions that verify Assumptions~\ref{as.str} and \ref{as.smooth} only for $\x \in X \subset \mathbb{R}^n$. We explore this scenario in the second numerical example. An alternative setting in which Assumptions~\ref{as.str} and \ref{as.smooth} need not hold is if we restrict (project) the algorithms to a neighborhood of the optimal trajectory.  In this latter case, the convergence analysis becomes local only. 

\end{remark}
\vskip2mm





We start the convergence analysis by deriving an upper bound on the norm of the approximation error $\mathbold{\Delta}_k\in \reals^n$ of the first-order forward Euler integral in \eqref{GTT_prediction} (w.r.t. the continuous dynamics~\eqref{eq.dyn}). This error is sometimes referred to as the local truncation error~\cite{Quarteroni2000}. The error is defined as the difference between the predicted $\x_{k+1|k}$ in~\eqref{GTT_prediction} and the exact prediction $\x(t_{k+1})$ obtained by integrating the continuous dynamics~\eqref{eq.dyn} from the same initial condition $\x_k$, i.e.,
\begin{align}\label{eq.def.delta}
\mathbold{\Delta}_k : = \x_{k+1|k} - \x(t_{k+1}) .
\end{align}
The upper bound for the norm $\|\mathbold{\Delta}_k \|$ is central in all our algorithms, since it encodes the error coming from the prediction step. We study this upper bound in the following proposition.

\vspace{1mm}
\begin{proposition}\label{prop.err}
Under Assumptions~\ref{as.str}-\ref{as.smooth}, the error norm $\|\mathbold{\Delta}_k\|$ of the Euler approximation~\eqref{GTT_prediction} defined in~\eqref{eq.def.delta} is upper bounded by
\begin{equation}\label{prop_claim_err_bound}
\|\mathbold{\Delta}_k\| \leq \frac{h^2}{2} \left[\frac{C_0^2 C_1}{m^3} + \frac{2 C_0 C_2}{m^2} + \frac{C_3}{m}\right] = O(h^2).
\end{equation}
\end{proposition}

\vskip1mm

\begin{IEEEproof}
See Appendix~\ref{ap.prop}.
\end{IEEEproof}

\vskip1mm

Proposition~\ref{prop.err} states that the norm of the discretization error $\|\mathbold{\Delta}_k\|$ is bounded above by a constant which is in the order of $O(h^2)$. We use this upper bound in proving convergence of all the proposed methods.

\subsection{Gradient trajectory tracking convergence}

We study the convergence properties of the sequence of variables $\x_k$ generated by GTT for different choices of the stepsize. In the following theorem we show that the optimality gap $\|\x_k-\x^*(t_{k})\|$ converges exponentially to an error bound.

\vskip1mm
\begin{theorem}\label{th.GTT_convg} 
Consider the gradient trajectory tracking algorithm as defined in \eqref{equivalent_problem}-\eqref{GTT_correction}. Let Assumptions~\ref{as.str}-\ref{as.smooth} hold true and define the constants $\rho$ and $\sigma$ as
\begin{equation}\label{thm:rho_sigma}
\rho := \max\{|1-\gamma m|,|1 - \gamma L|\}, \ \sigma := 1 + h(C_0 C_1/m^2 + C_2/m).
\end{equation}
Let the stepsize $\gamma$ be chosen as $0 <\gamma < 2/L$, which implies $\rho < 1$. 
\vskip-2mm
\begin{enumerate}[i)]
\item For any sampling period $h$, the sequence $\{\x_k\}$ converges to $\x^*(t_k)$ exponentially up to a bounded error as
\begin{align}\label{result2}
&\|{\x}_{k} - \x^*(t_{k})\| \leq \rho^{\tau k} \|{\x}_{0} - \x^*(t_{0})\| 
\\ &
\!\!\!\!\!\!
+\!  \rho^\tau \!\left[ h\left[\frac{2 C_0 }{m}\right] \!+\!\frac{h^2}{2}\! \left[\frac{C_0^2 C_1}{m^3} + \frac{2 C_0 C_2}{m^2} + \frac{C_3}{m}\right]\right]\!\left[ \frac{1-\rho^{\tau  k}}{1-\rho^\tau}\right].\nonumber
\end{align}
\item If the sampling period $h$ is chosen such that $\rho^\tau \sigma < 1$, i.e.,
\begin{equation}\label{res_cond_3}
h < \left[\frac{C_0 C_1}{m^2} + \frac{C_2}{m} \right]^{-1} \left({\rho^{-\tau}} -1 \right),
\end{equation}
then the sequence $\{\x_k\}$ converges to $\x^*(t_k)$ exponentially up to a bounded error as
\begin{align}\label{result1}
\|{\x}_{k} - \x^*(t_{k})\| &\leq (\rho^\tau \sigma)^{k} \|{\x}_{0} - \x^*(t_{0})\| 
\\ &
\!\!\!\!\!\!\!\!\!\!\!\!\!\!\!\!\!\!\!\!\!\!
+   \rho^\tau \frac{h^2}{2}\! \left[\frac{C_0^2 C_1}{m^3} + \frac{2 C_0 C_2}{m^2} + \frac{C_3}{m}\right] \left[ \frac{1-(\rho^\tau\sigma)^k}{1-\rho^\tau\sigma}\right].\nonumber
\end{align}
\end{enumerate} 
\end{theorem}

\vskip1mm

\begin{proof}
See Appendix \ref{app.GTT_convg}.
\end{proof}
\vskip1mm

Theorem \ref{th.GTT_convg} states the convergence properties of the GTT algorithm for different choices of the parameters. In both cases the exponential convergence to a neighborhood is shown, however, the accuracy of convergence depends on the choice of the sampling period $h$, the stepsize parameter $\gamma$, and the number of  gradient descent steps $\tau$. To guarantee that the constant $\rho$ is strictly smaller than $1$, the stepsize must satisfy $\gamma<2/L$ : this can be seen by the definition of $\rho$ and the fact that $m \leq L$ by Assumptions~\ref{as.str} and~\ref{as.smooth}. Then, for any choice of the sampling period $h$ the result in \eqref{result2} holds, which implies exponential convergence to a neighborhood of the optimal solution. In this case the error bound contains two terms that are proportional to $h$ and $h^2$. Therefore, we can say that the accuracy of convergence is in the order of $O(h)$. Notice that increasing the number of  gradient descent iterations $\tau$ improves the speed of exponential convergence by decreasing the factor $\rho^\tau$. Moreover, a larger choice of $\tau$ leads to a better accuracy since the asymptotic error bound is proportional to $\rho^\tau/(1-\rho^\tau)$.

The result in \eqref{result1} shows that the accuracy of convergence is proportional to the square of the sampling period $h$, if the sampling period is chosen to satisfy the condition $\rho^\tau \sigma < 1$. 
In the following corollary we formalize this observation by studying the asymptotic convergence results of GTT for different choices of stepsize.

\begin{corollary}\label{co.pcs}
Under the same conditions of Theorem \ref{th.GTT_convg}, the sequence of variables $\{\x_k\}$ generated by GTT converges to a neighborhood of $\x^*(t_k)$ asymptotically. The error bound when the parameters $\rho$ and $\sigma$ in \eqref{thm:rho_sigma} are chosen as $\rho^\tau \sigma \geq 1, \rho^\tau < 1$ is 
\begin{equation}\label{asy_1}
\limsup_{k\to\infty} \|\x(t_k) - \x^*(t_k)\| \leq  \frac{2 C_0 \rho^\tau h}{m(1-\rho^\tau)} = O(h) ,
\end{equation}
and if they satisfy $\rho^\tau \sigma < 1$ the error bound is
\begin{align}\label{asy_2}
&\limsup_{k\to\infty} \|\x(t_k) - \x^*(t_k)\| \nonumber\\ 
&\leq \frac{\rho^\tau h^2}{2(1-\rho^\tau\sigma)}\!\left(\frac{C_0^2 C_1}{m^3} + \frac{2 C_0 C_2}{m^2} + \frac{C_3}{m}\right)=O(h^2).
\end{align}

\end{corollary}
\vskip2mm

%

The asymptotic results in Corollary \ref{co.pcs} are implied by considering the results in Theorem \ref{th.GTT_convg} when  $k \to \infty$. Notice that when the stepsize satisfies conditions $\rho^\tau \sigma \geq 1, \rho^\tau < 1$ the convergence accuracy of GTT is in the order of $O(h)$. Moreover, if the sampling period $h$ is chosen such that $\rho^\tau \sigma<1$ then the error bound is in the order of $O(h^2)$.


\subsection{Newton trajectory tracking convergence}

Notice that the GTT algorithm does not incorporate the second-order information of the objective function $f(\x;t_{k+1})$ to correct the predicted variable $\x_{k+1|k}$, while the NTT algorithm uses Newton's method in the correction step. Similar to the advantages of Newton's method relative to the gradient descent algorithm, we expect to observe faster convergence and more accurate estimation for NTT relative to GTT. In particular, one would expect that if Newton's method is in its quadratic phase, the error should be at least in the order of $O(h^4)$. In the following theorem we show that when both the initial estimate $\x_0$ is close enough to the initial solution $\x^*(t_0)$ and the sampling period $h$ is chosen properly, then NTT yields a more accurate convergence relative to GTT. 


\vskip1mm
\begin{theorem}\label{th.NTT_convg} 
Consider the NTT algorithm generated by \eqref{GTT_prediction} and \eqref{NTT_correction}. Assume that all the conditions in Assumptions~\ref{as.str}-\ref{as.smooth} hold. Define constants $\delta_1$, $\delta_2$ and $Q$ as
\begin{equation}\label{eq.simply}
\delta_1 := \frac{C_0C_1}{m^2} + \frac{C_2}{m},\, \delta_2 := \frac{C_0^2 C_1}{2 m^3} + \frac{C_0 C_2}{m^2} + \frac{C_3}{2 m},  \, Q:=\frac{2m}{C_1}.
\end{equation}
Further, recall $\tau$ as the number of Newton steps in the correction step. For any constant $c>0$, if the sampling period $h$ satisfies
\begin{equation}\label{up2}
 h \leq \min\left\{1, \left[\frac{Q^{2\tau-1} c}{((1+\delta_1)c + \delta_2)^{2\tau}}\right]^{\frac{1}{4\tau-2}} \right\},
\end{equation}
and the initial error $ \|\x_0 - \x^*(t_0)\| $ satisfies the condition
\begin{align}\label{locality}
 \|\x_0 - \x^*(t_0)\| \leq c h^2 ,
\end{align}
then the sequence $\|\x_k - \x^*(t_k)\| $ generated by NTT for $k\geq1$ is bounded above as
\begin{equation}\label{NewtonResult2}
 \|\x_k - \x^*(t_k)\| \leq Q^{-(2\tau-1)}(\sigma c + \delta_2)^{2\tau}h^{4\tau}.
\end{equation} 
\end{theorem}

\vskip1mm

\begin{proof}
See Appendix~\ref{ap.newton}. 
\end{proof}
\vskip1mm

Theorem~\ref{th.NTT_convg} establishes that, under additional conditions, the NTT tracks the optimal trajectory $\x^*(t_k)$ up to an error bound not larger than 
\begin{equation}
Q^{-(2\tau-1)}(\sigma c+ \delta_2 )^{2\tau}h^{4\tau} = O(h^{4\tau}),
\end{equation}
where $h$ is the sampling period. This is a result of the quadratic convergence of Newton's method.

The conditions can be intuitively explained as follows. Condition~\eqref{locality} formalizes the local nature of the convergence analysis of Theorem~\ref{th.NTT_convg}: due to the dependence of~\eqref{up2} on $c$, the right-hand side of~\eqref{locality} is in fact upper bounded.  For example, when $c\to \infty$, then $h \to 0$ and $ch^2 \to Q/(1+\delta_1)$. We notice that the initial gap is proportional to $h^2$, since the integration error $\|\mathbold{\Delta}\|$ has the same dependence on $h$. 
Finally, \eqref{up2} derives an upper bound on the allowable sampling period. It comprises of two terms, the first coming from the need for a local analysis, the second from convergence arguments. 
%
Despite the fact that Theorem~\ref{th.NTT_convg} is a local convergence result, in the numerical simulations we will display how NTT behaves very well even in a global sense, and for $\tau = 1$ achieves the proven $O(h^4)$ error bound.


\begin{remark}\label{rem.quad}
\emph{(Quadratic functions and backtracking)} Conditions~\eqref{locality} is a locality requirement, which is rather typical in for the analysis of Newton methods. The closer the function $f(\x; t)$ is to be quadratic, the smaller the parameter $C_1$ is. When the function is quadratic, then $C_1 = 0$, which in turns means $Q, ch^2 \to \infty$, i.e., global convergence is achieved (as expected). When $C_1$ becomes important, then one can think of initializing the Newton method with a backtracking strategy (as done often in practice), see~\cite{Boyd2004a}.     
\end{remark}

\begin{remark}\label{remark.hybrid}
\emph{(Hybrid strategy)} Theorem~\ref{th.NTT_convg} suggests also a warm start procedure to implement the NTT algorithm. In particular, consider the condition $\|\x_0 - \x^*(t_0)\| \leq c h^2$. Given the strong convexity assumption and the fact that the gradient vanishes at optimality, this condition is implied by the following sufficient condition
\begin{equation}\label{eq.aryan100}
\|\nabla_{\x}f(\x_0; t_0)\| \leq  m\, c h^2,
\end{equation}
which is easier to check in practice than condition~\eqref{locality} (since normally one does not have access to the optimizer $\x^*(t_0)$). In fact, one might implement a hybrid strategy, where at the beginning we run the GTT algorithm and then we switch to NTT when the condition in \eqref{eq.aryan100} is satisfied. In order to make sure that the GTT algorithm eventually arrives at an error $\|\x_k - \x^*(t_k)\| \leq c h^2$, we need to pick $c$ in a way that $ c h^2$ is strictly bigger than the asymptotical error of GTT in~\eqref{asy_2}. Therefore, we must choose $c$ as 
\begin{equation}\label{eq.crho}
c > \frac{\rho^\tau \delta_2}{1-\rho^\tau\sigma}.
\end{equation}
Hence, start with GTT and choose a sampling period $h$ that verifies~\eqref{up2} and switch to NTT when condition~\eqref{eq.aryan100} is satisfied. We will see how this strategy performs in the simulation results. 
\end{remark}




\subsection{Convergence of methods with approximated time derivative }


We focus now on the approximated version of GTT and NTT (i.e., the AGT and ANT algorithms), where we approximate the time derivative of the gradient. In the following theorems, we formalize the fact that this approximation does not affect the order of the asymptotic error w.r.t. $h$. 

\vskip1mm
\begin{theorem}\label{th.AGT_convg} 
Consider the AGT algorithm as defined in Algorithm \ref{algo_AGT}, recall the definitions of the constants $\rho$ and $\sigma$ in \eqref{thm:rho_sigma}, and let Assumptions~\ref{as.str}-\ref{as.smooth} hold true. Let the stepsize $\gamma$ be chosen as $0 <\gamma < 2/L$, which implies $\rho < 1$.
\begin{enumerate}[i)]
\item For any sampling period $h$, the sequence $\{\x_k\}$ converges to $\x^*(t_k)$ exponentially up to a bounded error as
\begin{align}\label{result2a}
&\|{\x}_{k} - \x^*(t_{k})\| \leq \rho^{\tau k} \|{\x}_{0} - \x^*(t_{0})\| 
\\ &
\!\!\!\!\!\!\!
+\!  \rho^\tau \!\left[ h\left[\frac{2 C_0 }{m}\right] \!+\!\frac{h^2}{2}\! \left[\frac{C_0^2 C_1}{m^3} + \frac{2 C_0 C_2}{m^2} + \frac{2 C_3}{m}\right]\right]\!\left[ \frac{1-\rho^{\tau  k}}{1-\rho^\tau}\right].\nonumber
\end{align}
\item If the sampling period $h$ is chosen such that $\rho^\tau \sigma < 1$, i.e.,
\begin{equation}\label{res_cond_bully}
h < \left[\frac{C_0 C_1}{m^2} + \frac{C_2}{m} \right]^{-1} \left({\rho^{-\tau}} -1 \right),
\end{equation}
then the sequence $\{\x_k\}$ converges to $\x^*(t_k)$ exponentially up to a bounded error as,
\begin{align}\label{result1a}
\|{\x}_{k} - \x^*(t_{k})\| &\leq (\rho^\tau \sigma)^{k} \|{\x}_{0} - \x^*(t_{0})\| 
\\ &
\!\!\!\!\!\!\!\!\!\!\!\!\!\!\!\!\!\!\!\!\!\!\!\!\!\!\!\!\!
+   \rho^\tau \frac{h^2}{2}\! \left[\frac{C_0^2 C_1}{m^3} + \frac{2 C_0 C_2}{m^2} + \frac{2 C_3}{m}\right]  \left[ \frac{1-(\rho^\tau\sigma)^k}{1-\rho^\tau\sigma}\right].\nonumber
\end{align}
\end{enumerate} 
\end{theorem}

\vskip1mm

\begin{proof}
See Appendix \ref{ap.agt}.
\end{proof}
\vskip1mm
%

Theorem \ref{th.AGT_convg} states the convergence properties of the AGT algorithm for different choices of the parameters. In both cases the exponential convergence to a neighborhood is shown with convergence accuracy depending on the sampling period $h$, the stepsize $\gamma$, and the number of  gradient descent steps $\tau$. Moreover, for particular sampling period selections depending on smoothness properties of the objective, the asymptotic error bound either converges up to an $O(h)$ or $O(h^2)$ term. Notice that the convergence properties of AGT in \eqref{result2a} and \eqref{result1a} are identical to the convergence results of GTT in \eqref{result2} and \eqref{result1}, respectively, except for the coefficients of $h^2$. To be more precise, the coefficient of $h^2$ in \eqref{result2a} and \eqref{result1a} is ${C_0^2 C_1}/{2m^3} + {C_0 C_2}/{m^2} + {C_3}/{m}$, while the coefficient of $h^2$ in \eqref{result2} and \eqref{result1} is ${C_0^2 C_1}/{2m^3} +{C_0 C_2}/{m^2} + {C_3}/{2m}$. This observation implies that the error bound of AGT is slightly larger than the error of GTT which is implied by the error of the derivative approximation. However, the orders of the error bounds for these two algorithms are identical. 

AGT uses only first-order information of the objective $f(\x;t_{k+1})$ to correct the predicted variable $\x_{k+1|k}$, while ANT  uses the Newton method in the correction step. Similar to the advantages of NTT relative to GTT, we show more accurate estimation for ANT relative to AGT in the following theorem.

\vskip1mm
\begin{theorem}\label{th.ANT_convg} 

Consider the ANT algorithm as defined in Algorithm \ref{algo_ANT}, recall the definitions of the constants $\delta_1$, $\delta_2$ and $Q$ as in \eqref{eq.simply}, and let Assumptions~\ref{as.str}-\ref{as.smooth} hold true. Further, recall $\tau$ as the number of Newton steps in the correction step and define $\delta_2'$ as 
\begin{equation}\label{zahra}
\delta_2' := \delta_2 + \frac{C_3}{2m}.
\end{equation}
For any constant $c>0$, if the sampling period $h$ satisfies
\begin{equation}\label{up22200}
 h \leq \min\left\{1, \left[\frac{Q^{2\tau-1} c}{((1+\delta_1)c + \delta_2')^{2\tau}}\right]^{\frac{1}{4\tau-2}} \right\},
\end{equation}
and the initial error $ \|\x_0 - \x^*(t_0)\| $ satisfies the condition
\begin{align}\label{locality9090}
 \|\x_0 - \x^*(t_0)\| \leq c h^2 ,
\end{align}
then the sequence $\|\x_k - \x^*(t_k)\| $ generated by ANT for $k\geq1$ is bounded above as
\begin{equation}\label{NewtonResult222}
 \|\x_k - \x^*(t_k)\| \leq Q^{-(2\tau-1)}(\sigma c + \delta_2')^{2\tau}h^{4\tau}.
\end{equation} 
\end{theorem}
\vskip1mm

\begin{proof}
See Appendix \ref{ap.ant}.
\end{proof}
\vskip1mm

%

 Theorem~\ref{th.ANT_convg} states that the ANT algorithm reaches an estimation error of order $O(h^{4\tau})$. Observe that the error bound in \eqref{NewtonResult222} for ANT is slightly worse than the bound in \eqref{NewtonResult2} for NTT, since $\delta_2'>\delta_2$. On the other hand, the bound for both algorithms is in the order of $O(h^{4\tau})$. According to the results in Theorems~\ref{th.AGT_convg} and \ref{th.ANT_convg}, we can approximate the time derivative simply by a first-order scheme without changing the functional dependence of the error in $h$, but increasing its magnitude. In the simulation results, we show that this increase in error is in fact extremely limited. These analytical results therefore suggest the advantage of the proposed prediction-correction algorithms even in cases in which the knowledge of the time variability of the objective function is only estimated, which is important in many practical scenarios, e.g., in robotics or in statistical signal processing.    


\section{Numerical Experiments}\label{sec:num}

In this section, we implement the algorithms derived in Section \ref{sec:probform} for a couple practical examples in order to asses their performance in practice. Specifically, in Section \ref{sec:sims}, we consider a simple time-varying function and apply the GTT, NTT, AGT, and the hybrid method of Remark~\ref{remark.hybrid}. Additionally, in Section \ref{sec:robots}, we consider the task of designing a derivative control law for an autonomous system to follow a reference path. In this practical setting, we only consider the case where the time-derivative of the objective is not available, and hence must be approximated. Here this approximation corresponds to not having perfect information regarding the reference path the system aims to track.

\subsection{Scalar example}\label{sec:sims}

As a simple example, consider the case where the decision variable $x\in\reals$ is a scalar and the time-varying optimization problem is 
\begin{equation}\label{eq:scalar}
\min_{x \in \mathbb{R}} f(x; t): = \frac{1}{2}\left(x - \cos(\omega t)\right)^2 + \kappa \log[1 + \exp(\mu x)]. 
\end{equation} 
The function in \eqref{eq:scalar} represents, for instance, the goal of staying close to a periodically varying trajectory plus a logistic term that penalizes large values of $x$. The terms $\omega$, $\kappa$, and $\mu$ are arbitrary nonnegative scalar parameters. In our experiments these parameters are set to $\omega = 0.02\, \pi$, $\kappa = 7.5$, and $\mu = 1.75$. 
The function $f(x; t)$  satisfies all the conditions in Assumptions \ref{as.str} and \ref{as.smooth}. In particular, one can compute in close-form the quantities
\begin{subequations}
\begin{eqnarray}
\nabla_{xx} f(x; t) &&= 1 + \kappa \mu^2\, \textstyle\frac{\exp(\mu x)}{[1 + \exp(\mu x)]^2} , \\
\nabla_{tx} f(x; t) &&= \omega \sin(\omega t), \\
\nabla_{xxx} f(x; t) && = \kappa \mu^3\, \textstyle\frac{\exp(\mu x) [1- \exp(\mu x)]}{[1+ \exp(\mu x)]^3}, \\
\nabla_{xtx} f(x; t)  &&= 0.\\
\nabla_{ttx} f(x; t) &&= \omega^2 \cos(\omega t), 
\end{eqnarray}
\end{subequations}
and the bounds 
\begin{subequations}
\begin{eqnarray}
&& \hskip-1.3cm m = \min_{x\in\mathbb{R}, t} \nabla_{xx} f(x; t) = 1,\\
&& \hskip-1.3cm {L = \max_{x\in\mathbb{R}, t} \nabla_{xx} f(x; t)  = 1\!+\!\textstyle\frac{\kappa \mu^2}{4} = 6.7422} ,\\
&& \hskip-1.3cm C_0 = \max_{x\in\mathbb{R}, t}\nabla_{tx} f(x; t)  = \omega = 0.0628,\\ 
&& \hskip-1.3cm C_1\! =\!\max_{x\in\mathbb{R}, t}\nabla_{xxx} f(x; t)  \!=\! \kappa\mu^3\, \textstyle \frac{(2-\sqrt{3})(\sqrt{3}-1)}{[3+\sqrt{3}]^3} \!=\! 3.8678 , \\
&& \hskip-1.3cm C_2 = \max_{x\in\mathbb{R}, t} \nabla_{xtx} f(x; t)  = 0, \\ 
&& \hskip-1.3cm C_3 = \max_{x\in\mathbb{R}, t}
 \nabla_{ttx} f(x; t)   = \omega^2 = 0.0039. 
\end{eqnarray}
\end{subequations}

\begin{figure}[t]
\centering
{
\psfrag{x}[c]{\footnotesize Sampling time $t_k$}
\psfrag{y}[c]{\footnotesize Tracking error $\|x_k - x^*(t_k)\|$}
\psfrag{running gradient}{\footnotesize RG}
\psfrag{gradient}{\footnotesize GTT, $\tau = 1$}
\psfrag{approxgradient}{\footnotesize AGT, $\tau = 1$}
\psfrag{newton}{\footnotesize GTT, $\tau = 3$}
\psfrag{newton2}{\footnotesize GTT, $\tau = 5$}
\psfrag{newton3}{\footnotesize NTT, $\tau = 1$}
\psfrag{hybrid}{\footnotesize Hybrid,\! $\tau\! = \!1$}
\psfrag{bounds}{\footnotesize \hskip-.3cm Bounds Th.~1-3}
}
\includegraphics[width=0.45\textwidth]{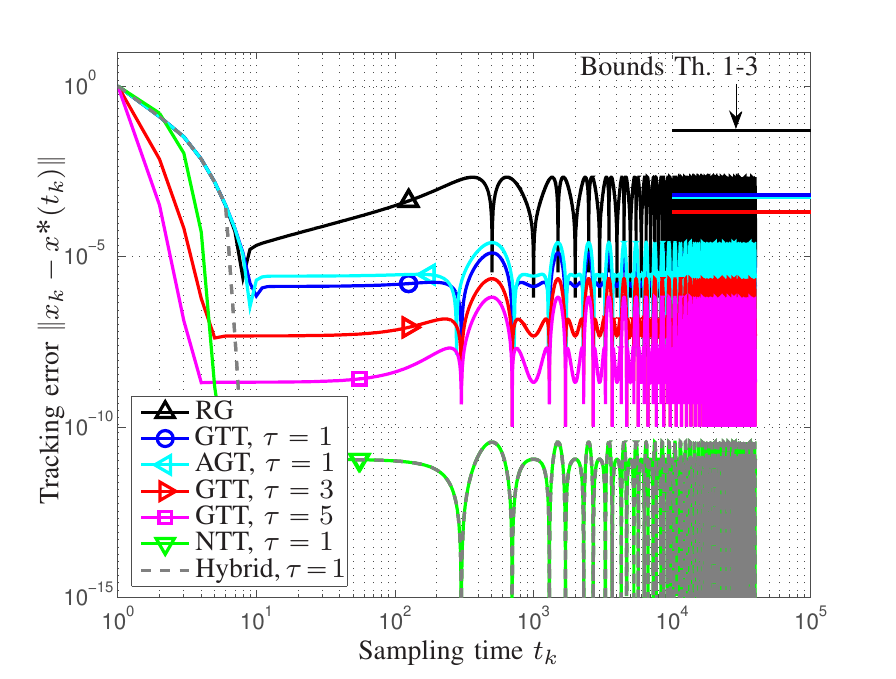}
\caption{Error with respect to the sampling time $t_k$ for different algorithms applied to the scalar problem~\eqref{eq:scalar}, with $h = 0.1$, $\kappa = 7.5$, $\mu = 1.5$.}
\label{fig.centr1}
\end{figure}
\begin{figure}[t]
\centering
{
\psfrag{x}[c]{\footnotesize Sampling period $h$}
\psfrag{y}[c]{\footnotesize Worst-case error}
\psfrag{running gradient}{\footnotesize RG}
\psfrag{gradient}{\footnotesize GTT, $\tau = 1$}
\psfrag{approxgradient}{\footnotesize AGT, $\tau = 1$}
\psfrag{newton}{\footnotesize GTT, $\tau = 3$}
\psfrag{newton2}{\footnotesize GTT, $\tau = 5$}
\psfrag{newton3}{\footnotesize NTT, $\tau = 1$}
\psfrag{oh}{\scriptsize $O(h)$}
\psfrag{oh2}{\scriptsize $O(h^2)$}
\psfrag{oh4}{\scriptsize $O(h^4)$}
}
\includegraphics[width=0.45\textwidth]{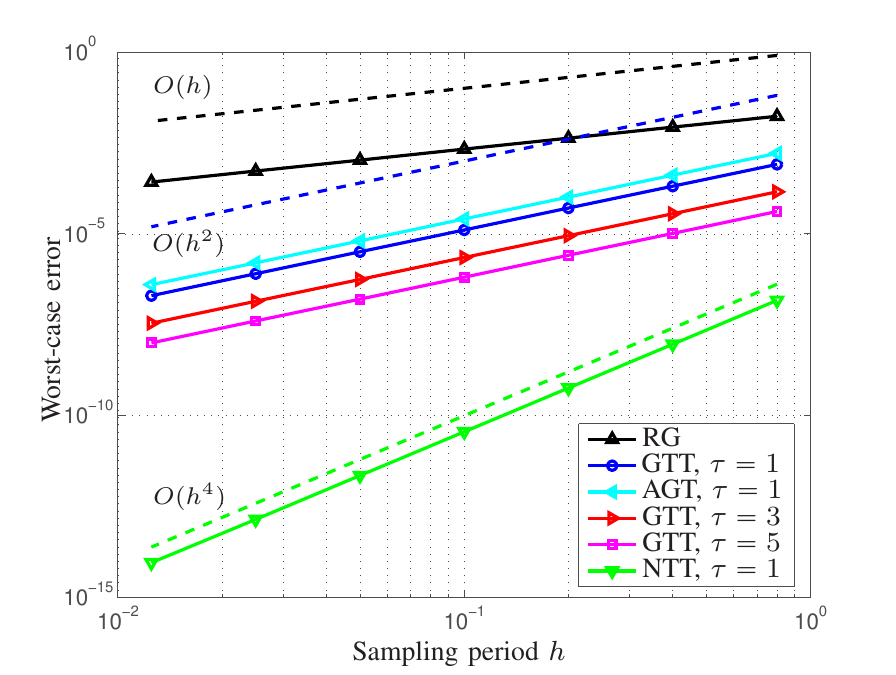}
\caption{Worst case error floor with respect to the sampling time interval $h$ for different algorithms applied to the scalar problem~\eqref{eq:scalar}, $\kappa = 7.5$, $\mu = 1.5$.}
\label{fig.centr2}
\end{figure}

\begin{figure}[t]
\centering
{
\psfrag{x}[c]{\footnotesize Sampling time $t_k$}
\psfrag{y}[c]{\footnotesize Tracking error $\|x_k - x^*(t_k)\|$}
\psfrag{running gradient}{\footnotesize RG}
\psfrag{gradient}{\footnotesize GTT, $\tau = 1$}
\psfrag{approxgradient}{\footnotesize AGT, $\tau = 1$}
\psfrag{newton}{\footnotesize GTT, $\tau = 3$}
\psfrag{newton2}{\footnotesize GTT, $\tau = 5$}
\psfrag{newton3}{\footnotesize NTT, $\tau = 1$}
\psfrag{hybrid}{\footnotesize Hybrid,\! $\tau\! = \!1$}
\psfrag{bounds}{\footnotesize \hskip-.3cm Bounds Th.~1-3}
}
\includegraphics[width=0.45\textwidth]{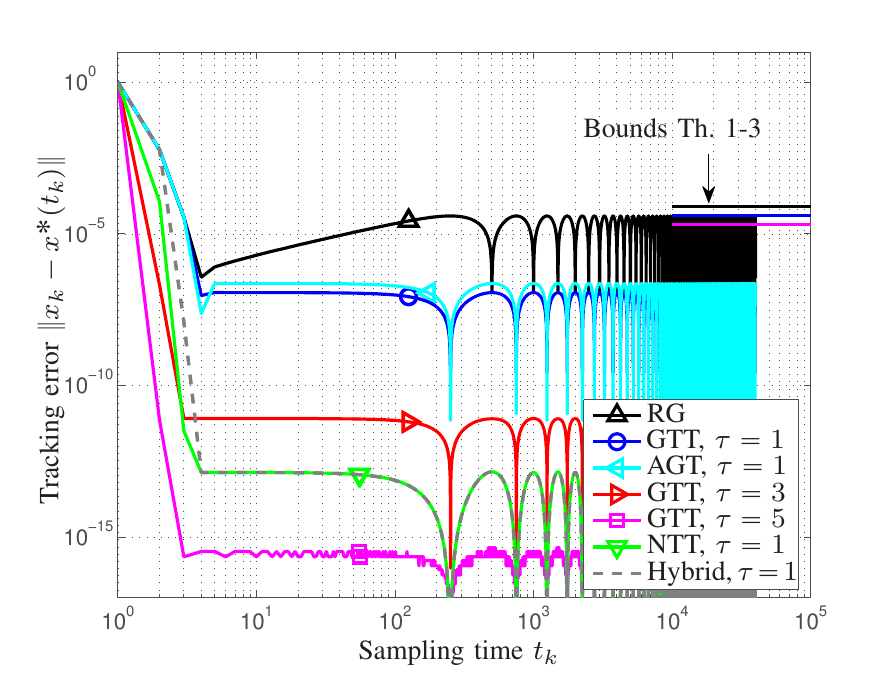}
\caption{Error with respect to the sampling time $t_k$ for different algorithms applied to the scalar problem~\eqref{eq:scalar}, with $h = 0.1$, $\kappa = .1$, $\mu = .5$}
\label{fig.centr12}
\end{figure}
\begin{figure}[t]
\centering
{
\psfrag{x}[c]{\footnotesize Sampling period $h$}
\psfrag{y}[c]{\footnotesize Worst-case error}
\psfrag{running gradient}{\footnotesize RG}
\psfrag{gradient}{\footnotesize GTT, $\tau = 1$}
\psfrag{approxgradient}{\footnotesize AGT, $\tau = 1$}
\psfrag{newton}{\footnotesize GTT, $\tau = 3$}
\psfrag{newton2}{\footnotesize GTT, $\tau = 5$}
\psfrag{newton3}{\footnotesize NTT, $\tau = 1$}
\psfrag{oh}{\scriptsize $O(h)$}
\psfrag{oh2}{\scriptsize $O(h^2)$}
\psfrag{oh4}{\scriptsize $O(h^4)$}
}
\includegraphics[width=0.45\textwidth, trim=0 0cm 0 0cm, clip=on]{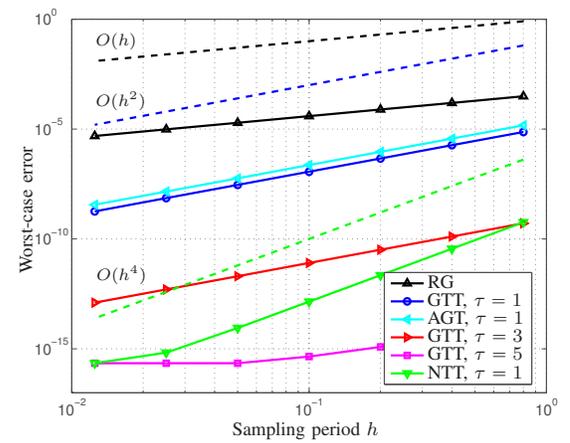}
\caption{Worst case error floor with respect to the sampling time interval $h$ for different algorithms applied to the scalar problem~\eqref{eq:scalar}, $\kappa = .1$, $\mu = .5$.}
\label{fig.centr22}
\end{figure}


We choose the constant stepsize as $\gamma = 0.2< 2/L$ in the gradient method stated in \eqref{GTT_correction} and initialize $x_0=0$ for all the algorithms. According to~\eqref{res_cond_3} the sampling period that guarantees an $O(h^2)$ error bound needs to be chosen as $h < 1.028.$ for all $\tau\geq 1$.

In Figure~\ref{fig.centr1}, we plot the error $\|x_k - x^*(t_k)\|$ versus the discrete time $t_k$ for a sampling period of $h = 0.1$, for different schemes, along with the asymptotical bounds computed via Theorems~\ref{th.GTT_convg} and \ref{th.AGT_convg}. Observe that the running gradient (RG) method~\cite{Popkov2005} which uses only a gradient correction step (and no prediction) performs the worst, achieving an error of $10^{-2}$, while GTT for $\tau=1$, $\tau=3$, and $\tau=5$ achieves an error of approximately $10^{-5}$. Numerically we may conclude that tracking with gradient-based prediction (GTT) for different values of $\tau$ has a better error performance than running, even in the case we use an approximate time derivative (AGT); in addition, tracking with Newton-based prediction (NTT) with $\tau=1$ achieves a superior performance compared to the others, i.e., an error stabilizing near $10^{-10}$ is achieved. 

In Figure~\ref{fig.centr1}, we also display the behavior of the hybrid strategy advocated in Remark~\ref{remark.hybrid}. We can see how after we switch to NTT (when the condition $\|\nabla_{\x}f(\x_k; t_k)\| \leq 0.0034$, derived from~\eqref{eq.crho}, is met), then in only one step we regain the same performance as NTT.

\begin{figure}[t]
\centering
{
\psfrag{x}[c]{\footnotesize Horizontal Position Coordinate $x$ [m] }
\psfrag{y}[c]{\footnotesize Vertical Position Coordinate $y$ [m] }
\psfrag{RG}{\footnotesize RG }
\psfrag{GTT2GTT2GTT2GTT2GTT1}{\footnotesize AGT, $\tau = 1$}
\psfrag{NTT}{\footnotesize ANT, $\tau = 1$}
\psfrag{True}{\footnotesize Optimal trajectory}
\psfrag{ObjectObject}{\footnotesize Reference path}
}
\includegraphics[width=.5\textwidth]{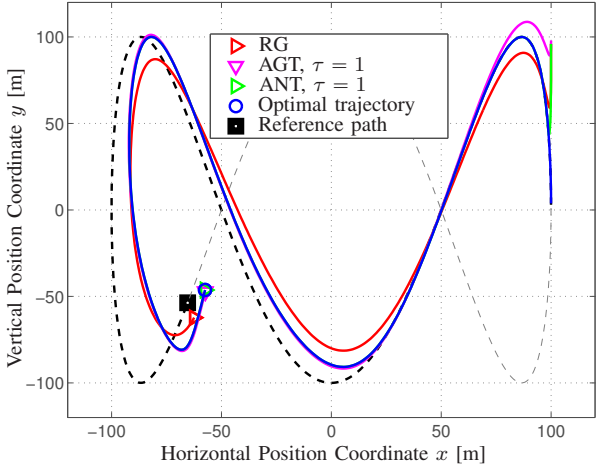}
\caption{Sample trajectories of the object to be tracked (dashed) and trajectories generated by the different algorithms (continuous). All algorithms track the optimum effectively, yet AGT and ANT track $\x^*(t)$ closer than RG.}\label{fig:trajectories}
\end{figure}


The differences in performance can be also appreciated by varying $h$ and observing the worst case error floor size which is defined as 
$ 
\max_{k>\bar{k}} \{\|x_k - x^*(t_k)\|\},
$
where $\bar{k} = 10^4$ in the simulations. Figure~\ref{fig.centr2} illustrates the error as a function of $h$. The performance differences between the proposed methods that may be observed here corroborate the differences evident in Figure \ref{fig.centr1}. In particular, the running method achieves the largest worst case error bound, followed in descending order by AGT, GTT with increasing $\tau$, and lastly NTT (or equivalently the hybrid strategy), which achieves the minimal worst-case error bound. Notice also the dashed lines displaying the theoretical performance of $O(h)$, $O(h^2)$, and $O(h^4)$, which are attained in this simulation. 

We continue the simulation example by changing the parameters $\kappa$ and $\mu$ in~\eqref{eq:scalar}  to the values $\kappa = .1$ and $\mu = 0.5$. This brings $L = 1.0063$, and a condition number $L/m$ close to $1$. In this settings a first-order method, such as the gradient, is expected to perform better than in the case of high condition numbers (as in the previous example). We pick the stepsize $\gamma = 1 < 2/L$. In Figures~\ref{fig.centr12} and \ref{fig.centr22}, we appreciate how the relative performances of GTT and NTT change with the new parameters\footnote{The code of the simulation example will be made available for the readers, to appreciate how different stepsizes may influence the asymptotical bounds.}.

\subsection{Target Tracking Experiments}\label{sec:robots}

The second numerical example consists of a more realistic application scenario. We consider an autonomous system (i.e., a mobile robot) which is charged with the task of following an object whose position is varying continuously in time. Denote the reference trajectory of this object as a curve $\y(t)$, i.e. a function $\y : \reals_+ \rightarrow \reals^n$ and $\x\in\reals^n$ be the decision variable of the robot, in terms of the waypoint it aims to reach next.
We aim to solve tracking problems of the form
\begin{equation}\label{eq:robot_track}
\min_{\x\in \mathbb{R}^2} f(\x; t): = \,\frac{1}{2} \left( \|\x - {\y}(t)\|^2 + \mu_1 \exp(\mu_2 \|\x - \b\|^2)\right),
\end{equation}
which corresponds to tracking the reference path $\y(t)$ while remaining close enough to a base station located in $\b$, which may correspond to a recharging station or a domain constraint associated with maintaining viable communications. Using the methods developed in Section \ref{sec:probform} for problems of this type correspond to deriving derivative-based control laws for fully actuated systems with simple integrator dynamics. 

For the example considered here, we consider a planar example ($n=2$) and fix $\mu_1 = 1000$~m$^2$, $\mu_2 = .005$~m$^{-2}$ with the base located at $\b = [100; 100]$~m. In addition, we suppose the target trajectory ${\y}(t)$ follows the specified path
$$
{\y}(t) = 100 [\cos(\omega t), \sin(3 \omega t)]~\textrm{m}
$$
where $\omega = 0.01$~Hz. Moreover, the position domain is given as $X = [-150,150]\times[-150,150]$~m$^2$ and we know that $\x^*(t)\in X$. We can compute the constants of Assumptions~\ref{as.str} and \ref{as.smooth} over $X \subset \mathbb{R}^n$ [Cfr. Remark~\ref{remark.as}] %
$
m = 1.01,  L = 3.45,  C_0 = 3.16~[\textrm{m/s}],
$
$
C_1 = 0.06~[\textrm{m}^{-1}],  C_2 = 0,   C_3 = 0.10~[\textrm{m/s}^2].
$
We select stepsize $\gamma = 0.05<2/L$. With these parameters and $h\!=\!1$s, the target moves with maximum speed of  $3.16$~m/s. This is comparable with the speed of current quad-rotors (max speed $\sim\hskip-0.1cm 10$~m/s). 
%

In any practical setting,  the actuation capability of an autonomous system is limited either in terms of velocity or degrees of freedom. We consider the case where the autonomous system may move with the same number of degrees as its decision variable dimension, i.e. it may move in any direction, yet its maximum velocity is limited to some value $v_{\max}$. A typical velocity maximum for ground vehicles is $v_{\max}=4$~m/s, which is the choice made in the numerical experiments here. Thus, we modify our algorithms to account for this constraint by rescaling the prediction-correction step to the allowable velocity limit. Of course more complicated actuation models may be considered, but these are beyond the scope of this work.

%
\begin{figure}[t]
\centering
{
\psfrag{x}[c]{\footnotesize Sampling time $t_k$ [s] }
\psfrag{y}[c]{\footnotesize Tracking Error $\|\x_k - \x^*(t_k)\|$ [m] }
\psfrag{RG}{\footnotesize RG }
\psfrag{GTT1}{\footnotesize AGT, $\tau = 1$}
\psfrag{GTT2GTT2GTT2GTT2}{\footnotesize AGT, $\tau = 3$}
\psfrag{GTT3}{\footnotesize AGT, $\tau = 5$}
\psfrag{NTT}{\footnotesize ANT, $\tau = 1$}
}
\includegraphics[width=.5\textwidth]{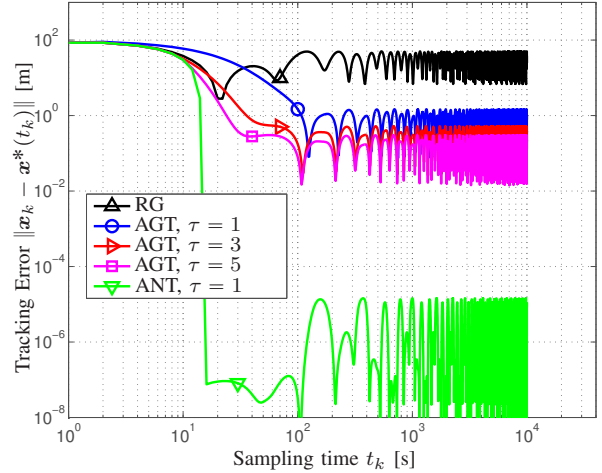}
\caption{Error~[m] with respect to the sampling time $t_k$ for $h = 1$~[s] for different algorithms applied to the tracking problem~\eqref{eq:robot_track}.}\label{fig:error_t}
\end{figure}


We show the result of this experiment in terms of the actual reference path and trajectories generated by the approximate algorithms AGT and ANT in Figure \ref{fig:trajectories} over a truncated time interval $0<t<300$~s. The reference trajectory $\y(t)$ is the dotted line, and the optimal continuous-time trajectory $\x^*(t)$ associated with solving \eqref{eq:robot_track} is in blue. By running gradient we mean a method which has no prediction step, and operates only by correction. Observe that the trajectories generated running gradient (RG), AGT, and ANT successfully track the optimal trajectory $\x^*(t)$, and consequently the reference path $\y(t)$ up to a small error. 

This trend may be more easily observed in Figure \ref{fig:error_t} which shows the magnitude of the difference between the generated path and the optimal path $\| \x^*(t_k) - \x_k\|$, or the tracking error, as compared with the sampling time $t_k$. Note that the asymptotical bounds computed via Theorems~\ref{th.GTT_convg} and \ref{th.AGT_convg} are less meaningful here since the velocity of the robot is scaled. The approximate steady state errors achieved by RG, AGT, and ANT are respectively $10$, $10^{-1}$, and $10^{-5}$. AGT experiences comparable levels of error across different values of $\tau$, the number of correction steps, and ANT far outperforms the other methods.
This pattern is corroborated in Figure \ref{fig:pessistic_err}, which plots the worst-case error $ \max_{k \geq \bar{k}}\| \x^*(t_k) - \x_k\|$ versus the sampling interval size $h$ for $\bar{k}=8\times10^3$. In particular, we observe that RG experiences an error comparable to $O(h)$, as it theoretically guarantees, whereas our proposed methods AGT and ANT achieve a worst-case error of approximately $O(h^2)$ and $O(h^4)$, respectively. Observe that as the problem \eqref{eq:robot_track} is sampled less often, i.e. when $h$ increases, the optimality gap increases.
\begin{figure}[t]
\centering
{
\psfrag{x}[c]{\footnotesize Sampling period $h$ [s] }
\psfrag{y}[c]{\footnotesize Worst-case error [m] }
\psfrag{RG}{\footnotesize RG }
\psfrag{GTT1}{\footnotesize AGT, $\tau = 1$}
\psfrag{GTT2GTT2GTT2GTT2}{\footnotesize AGT, $\tau = 3$}
\psfrag{GTT3}{\footnotesize AGT, $\tau = 5$}
\psfrag{NTT}{\footnotesize ANT, $\tau = 1$}
\psfrag{oh}{\footnotesize $O(h)$}
\psfrag{oh2}{\footnotesize $O(h^2)$}
\psfrag{oh3}{\footnotesize $O(h^4)$}
}
\includegraphics[width=.5\textwidth]{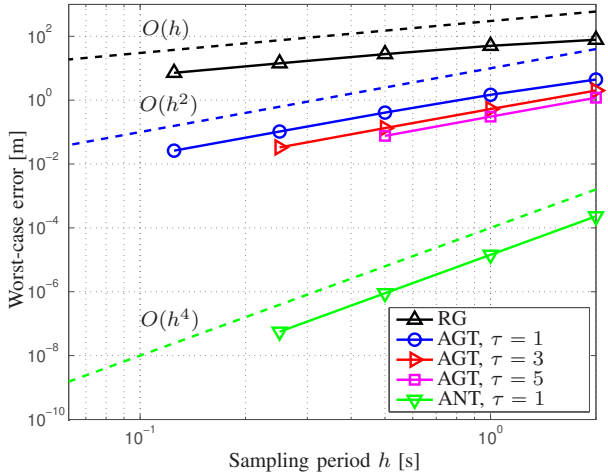}
\caption{Worst case error floor with respect to the sampling time interval $h$ for different algorithms applied to the tracking problem~\eqref{eq:robot_track}.}
\label{fig:pessistic_err}
\end{figure}

{\it Computational Considerations.} 
We empirically observe ANT to far outperform the other methods; however, this performance gap ignores the increased computational cost associated with Newton steps. To obtain a more fair comparison, we consider how the different algorithms perform when the computational time per correction and prediction steps are fixed. Theoretically, each prediction step and Newton step require $O(n^3)$ computations (because of the matrix inversion), while the gradient step only $O(n)$. Practically, in this simulation setting, the most demanding task is however the \emph{evaluation} of the gradient and the Hessian, while the actual prediction or correction step is less critical (less than 1/10 time). In particular, evaluating the Hessian requires \emph{twice} the computational effort of evaluating the gradient, so a Newton step is three times slower than a gradient step.

The workflow for each optimization iteration is the following:
\vskip2mm

\fbox{\begin{minipage}{0.44\textwidth}
\begin{enumerate}
\item[$t_k$)] A new function is acquired;
\item[1)] A new way point $\x_{k}$ is generated via a correction step;
\item[2)] The way point is implemented and the robot moves;
\item[3)] Either a new prediction $\x_{k+1|k}$ is made, based on past information, or the correction is refined by more correction steps. 
\end{enumerate} 
\end{minipage}
}

\vskip2mm

We see that at step 3 the robot can either implement the prediction part of our prediction-correction algorithms, or refine the correction to have, perhaps, a better starting point when the next function is acquired. We consider here the running gradient RG (which we remark is nothing less than AGT without prediction), the AGT, the ANT, and a running version of the Newton method, which uses only correction steps (later indicated as RN). 
%

\begin{table}[t]
\centering
\caption{Number of correction steps to keep the same computational time}\label{table1}
\begin{tabular}{c|ccccccc}
Sampling period $h$~[s] & $1/10$ & $1/4$ & $1/3$ & $1/2$ & $2/3$ & $3/4$ & $1$ \\ \hline
RG & $1$  & $3$ & $4$ & $6$ & $8$ & $9$ & $12$ \\ 
RN & $-$  & $1$ & $1$ & $2$ & $2$ & $3$ & $4$ \\ 
AGT  & $1$  & $3$ & $4$ & $6$ & $8$ & $9$ & $12$ \\ 
ANT & $-$  & $1$ & $1$ & $2$ & $2$ & $3$ & $4$ \\
\end{tabular}
\end{table}

\begin{figure}[t]
\centering
{
\psfrag{x}[c]{\footnotesize Sampling period $h$ [s] }
\psfrag{y}[c]{\footnotesize Worst-case error [m] }
\psfrag{RG1}{\footnotesize  RG + 3G }
\psfrag{RG2}{\footnotesize  RG + 1N}
\psfrag{RN}{\footnotesize  RN + 1N}
\psfrag{AGT}{\footnotesize  AGT}
\psfrag{GTTGTT22}{\footnotesize  AGT}
\psfrag{GTT3}{\footnotesize  AGT, $\tau = 5$}
\psfrag{ANT}{\footnotesize  ANT}
}
\includegraphics[width=.5\textwidth]{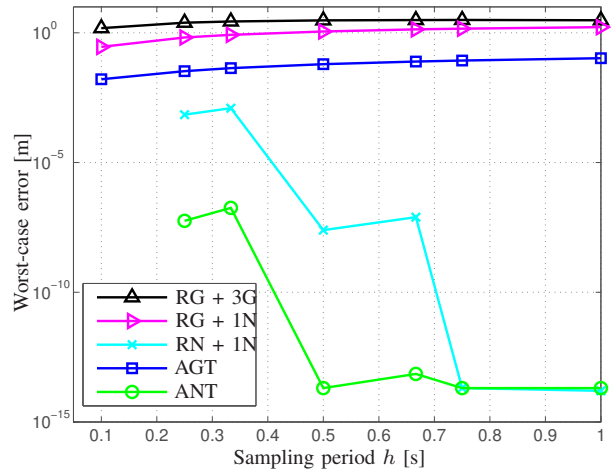}
\caption{Worst case Error~[m] w.r.t. $h$ [s] with fixed computational complexity.}\label{fig:complexity_fixed}
\end{figure}

We now outline how Table 1 is generated. We set $\Delta t_{\textrm{c}} =  h/10$ as the allowable computational time for the correction step (step 1), and we set the gradient evaluation to require $1/120$~s. As a consequence, for this setting the robot can perform only $\tau=1$ gradient correction step for a sampling time of $h = 0.1$~s. With this as our basic unit of measurement, we fill in Table 1 with how many gradient evaluations $\tau$ may be afforded with increasing the sampling interval $h$. As previously noted, ANT requires three times the computation time of AGT, and consequently experiences too much latency to be used when $h=.1$~s.

We set as $\Delta t_{\textrm{p}} =  1/40$~s as the allowable computational time for step 3, so that we can either run one prediction step, $3$ gradient correction refinement steps, or $1$ Newton correction refinement step. 

We run the different algorithms when the computation time is fixed (i.e. for $h=.1$~s, in step 1. $\tau=1$ steps of RG and AGT may be afforded, but {\it zero} of ANT) and record the worst-case error achieved versus $h$  in Figure \ref{fig:complexity_fixed}. We run RG both with $3$ additional gradient refinement steps (3G) and with $1$ Newton refinement step (1N), while RN is run with $1$ Newton refinement (1N). Broadly, one may observe that if ANT may be afforded (i.e. for large $h$), it is much preferable to AGT regardless of the number of correction steps $\tau$. However, for small sampling periods $h$, i.e. when one requires very low latencies in the control loop, ANT is infeasible. We also observe that prediction is to be preferred to additional refinement steps, especially when the sampling period is small (i.e., when the time derivative approximation makes a significant difference because one does not have enough time to perform many correction steps).

\section{Conclusion}\label{sec:conclusions}
We have designed algorithms to track the solution of time-varying unconstrained and strongly convex optimization problems. These algorithms leverage the knowledge of how the cost function changes in time and are based on a predictor-corrector scheme. We have also developed approximation schemes for when the rate at which the objective varies in time is not known. We established that these methods yield convergence to a neighborhood of the optimal trajectory, with a neighborhood of convergence dependent on the sampling period. Moreover, the size of this neighborhood is an order of magnitude smaller than state-of-the art running algorithms which only perform correction steps. In some cases when the problem parameters are appropriately chosen and second-order information is incorporated, the neighborhood of the optimal trajectory to which the algorithm converges is several orders of magnitude smaller than existing approaches. 

Moreover, we conducted a numerical analysis of the proposed methods in a simple setting which empirically supported the established error bounds. We also considered the task of developing a control strategy for an autonomous system to follow an object whose position varies continuously in time, showing that the developed tools yield an effective strategy. In some cases, the algorithms which achieve higher accuracy require\! too\! much\! computational latency to be used in a closed loop control setting; however, when this latency may be afforded, the second-order methods yield highly accurate tools.

Future research directions encompass the generalization of this work to constrained problems, general convex cost functions, as well as approximate second-order methods to weaken the computational requirements of computing the Hessian inverse in the prediction step.

%

\appendices


\section{Proof of Proposition~\ref{prop.err}}\label{ap.prop}

Let us analyze the forward Euler method applied to the vector-valued nonlinear dynamical system
\begin{equation}\label{proof_110}
\dot{\x} = {\mathbold{F}}(\x(t),t).
\end{equation}
If we apply the forward Euler method to the relation in \eqref{proof_110}, starting at a certain point $\x(t_k)$, we obtain 
\begin{equation}\label{proof_120}
\x_{k+1|k} = \x(t_k)+ h\, {\mathbold{F}}(\x(t_k),t_k). 
\end{equation}
On the other hand, we can write $\x(t_{k+1})$ by using a Taylor expansion as 
\begin{equation}\label{proof_120a}
\x(t_{k+1}) = \x(t_k)+ h\, {\mathbold{F}}(\x(t_k),t_k) + \frac{h^2}{2} \frac{\textrm d}{\textrm d t} \, {\mathbold{F}}(\x(s),s), 
\end{equation}
for a certain time $s \in [t_k, t_{k+1}]$. Subtracting $\x(t_{k+1})$ from the both sides of the equality in \eqref{proof_120} and computing the norm of the resulting relation implies that
\begin{equation}\label{proof_130} 
\|\x_{k+1|k} - \x(t_{k+1})\| \!= \! \left\| \frac{h^2}{2} \frac{\textrm d}{\textrm d t} \, {\mathbold{F}}(\x(s),s)  \right\|.
\end{equation}
By considering the definition of the discretization error vector $\mathbold{\Delta}_k\!:=\!\x_{k+1|k}\!-\!\x(t_{k+1})$, we can write \eqref{proof_130}~as
\begin{equation}\label{proof_160} 
\|\mathbold{\Delta}_k \|=  \frac{h^2}{2} \left\|\frac{\textrm d}{\textrm d t}{\mathbold{F}}(\x({s}),{s})\right\|.
\end{equation}
We proceed to find an upper bound for the right-hand side of \eqref{proof_160}. 
Observing the continuous dynamical system in \eqref{eq.dyn} we know that ${\mathbold{F}}(\x(t),t)$ is given by
\begin{equation}\label{proof_170} 
{\mathbold{F}}(\x(t),t) = -[\nabla_{\x\x}f(\x;t)]^{-1} \nabla_{t\x}f(\x;t).
\end{equation}
Then, by the chain rule we can write
\begin{align}\label{proof_170chain} 
\frac{\textrm d}{\textrm d t}{\mathbold{F}}(\x(t),t) &= \nabla_t {\mathbold{F}}(\x,t) + [\nabla_{\x} {\mathbold{F}}(\x,t)]\,\dot{\x} \nonumber \\
&=\nabla_t {\mathbold{F}}(\x,t) + [\nabla_{\x} {\mathbold{F}}(\x,t)]\,{\mathbold{F}}(\x(t),t),
\end{align}

where we have used the relation~\eqref{proof_110}. By using the triangle inequality, we can upper bound the norm of the right-hand side of~\eqref{proof_170chain} as
\begin{equation}\label{proof_170chain_b} 
\left\|\frac{\textrm d}{\textrm d t}{\mathbold{F}}(\x(t),t)\right\| \leq \| \nabla_t {\mathbold{F}}(\x,t) \| + \|[\nabla_{\x} {\mathbold{F}}(\x,t)]\,{\mathbold{F}}(\x(t),t)\|.
\end{equation}
We now upper bound the right-hand side of~\eqref{proof_170chain_b} by analyzing its two components. First, based on the definition in \eqref{proof_170}, the partial derivative w.r.t. time can be written as, $\nabla_t {\mathbold{F}}(\x,t) = $
$-\nabla_{t}\left[[\nabla_{\x\x}f(\x;t)]^{-1} \nabla_{t\x}f(\x;t)\right]$. By applying the chain rule, 
\begin{align}\label{proof_190} 
&\nabla_{t}\left[[\nabla_{\x\x}f(\x;t)]^{-1} \nabla_{t\x}f(\x;t)\right]\!=\!\![\nabla_{\x\x}f(\x; t)]^{-1}\nabla_{tt\x}f(\x; t)\nonumber\\
&\,\qquad\qquad\qquad-[\nabla_{\x\x}f(\x; t)]^{-2} \nabla_{t \x\x}f(\x; t) \nabla_{t\x}f(\x; t).
\end{align}
Compute the norm of both sides of \eqref{proof_190}. Substitute the norm $\|\nabla_{t}\left[[\nabla_{\x\x}f(\x;t)]^{-1} \nabla_{t\x}f(\x;t)\right]\|$ by $\|\nabla_{t}{\mathbold{F}}(\x,t) \|$. Further, apply the triangle inequality to the right-hand side of the resulting expression to obtain 
\begin{align}\label{proof_200} 
\|\nabla_{t}{\mathbold{F}}(\x,t) \|&\leq\left\|[\nabla_{\x\x}f(\x; t)]^{-2} \nabla_{t \x\x}f(\x; t) \nabla_{t\x}f(\x; t)\right\|\nonumber\\
&
\qquad+ \left\|[\nabla_{\x\x}f(\x; t)]^{-1}\nabla_{tt\x}f(\x; t)\right\|.
\end{align}
Observe the fact that $\nabla_{t \x\x}f(\x; t) = \nabla_{\x t\x}f(\x; t)$. We use the Cauchy-Schwartz inequality and the bounds in Assumptions \ref{as.str} and  \ref{as.smooth} to update the upper bound in \eqref{proof_200} as
\begin{equation}\label{proof_210} 
\|\nabla_{t}{\mathbold{F}}(\x,t) \|  \leq \frac{C_0 C_2}{m^2} + \frac{C_3}{m}.
\end{equation}
We can now do the same for the second component of the right-hand side of~\eqref{proof_170chain_b}, and in particular
\begin{align}\label{proof_220} 
\|\nabla_{\x}{\mathbold{F}}(\x,t)\,{\mathbold{F}}(\x(t),t) \| &= \|([\nabla_{\x\x}f(\x; t)]^{-1}\nabla_{\x t\x}f(\x; t)-\nonumber\\
& \hspace*{-3cm} [\nabla_{\x\x}f(\x; t)]^{-2} \nabla_{\x\x\x}f(\x; t) \nabla_{t\x}f(\x; t)) \,{\mathbold{F}}(\x(t),t) \| \nonumber \\ 
&\leq \left(\frac{C_2}{m} + \frac{C_1 C_0}{m^2}\right) \, \frac{C_0}{m}.
\end{align}
By combining the relation in \eqref{proof_160} and~\eqref{proof_170chain_b} with the upper bounds in \eqref{proof_210} and \eqref{proof_220}, the claim in \eqref{prop_claim_err_bound} follows.
\qed 

\section{Proof of Theorem \ref{th.GTT_convg}}\label{app.GTT_convg}

In order to prove Theorem \ref{th.GTT_convg}, we start by bounding the error in the prediction step by the terms that depend on the functional smoothness and the discretization error using Taylor expansions. Then we bound the tracking error of the gradient step using convergence properties of the gradient on strongly convex functions. By substituting the error of the correction step into the prediction step, we establish the main result.

%
%

First, we establish that discrete-time sampling error bound stated in \eqref{result1} is achieved by the updates \eqref{GTT_prediction}-\eqref{GTT_correction}. For simplicity, we modify the notation to omit the arguments $\x_k$ and $t_k$ of the function $f$. In particular, define 
\begin{align}\label{eq:thm1_defs}
&\nabla_{\x\x}f := \nabla_{\x\x}f(\x_k; t_k)\; , &&\nabla_{t\x} f := \nabla_{t\x} f(\x_k; t_k)\;, \\ 
&\nabla_{\x\x}f^* := \nabla_{\x\x}f({\x}^*(t_{k}); t_k)\;,  && \nabla_{t\x} f^* := \nabla_{t\x} f({\x}^*(t_{k}); t_{k}) \; . \nonumber
\end{align}
Begin by considering the update in~\eqref{GTT_prediction}, the prediction step, evaluated at a generic point $\x_k$ sampled at the current sample time $t_k$ and with associated optimizer $\x^*(t)$, which due to optimality will have null residual vector $\r(t) = \mathbf{0}$. Thus we may write
\begin{equation}\label{eq:null_residual}
\left\{\begin{array}{rcl}
{\x}_{k+1|k} &=& \x_{k} - h\,[\nabla_{\x\x}f]^{-1}\nabla_{t\x} f\\
{\x}^*(t_{k+1}) &=& {\x}^*(t_{k}) - h\, [\nabla_{\x\x}f^*]^{-1} \nabla_{t\x} f^* + \mathbold{\Delta}_k.
\end{array}\right.
\end{equation}
By subtracting the equalities in \eqref{eq:null_residual}, considering the norm of the resulting expression, and applying the triangle inequality we obtain
\begin{align}\label{eq.imp}
\|{\x}_{k+1|k}& -  \x^*(t_{k+1})\| \leq \|{\x}_{k} - \x^*(t_{k})\|  \\  & + h\left\|[\nabla_{\x\x}f]^{-1}\nabla_{t\x} f - [\nabla_{\x\x}f^*]^{-1}\nabla_{t\x} f^*  \right\|  + \|\mathbold{\Delta}_k\| .\nonumber 
\end{align}
Substituting the discretization error norm $\|\mathbold{\Delta}_k\|$ by its upper bound in \eqref{prop_claim_err_bound} follows
\begin{align}\label{eq.imp00}
\|{\x}_{k+1|k}&\! -\!  \x^*(t_{k+1})\| \!\leq\! \|{\x}_{k}\! - \!\x^*(t_{k})\| \! \nonumber \\ &\quad\quad+\! \frac{h^2}{2}\!\left[\!\frac{C_0^2 C_1}{m^3}\!  +\! \frac{2 C_0 C_2}{m^2}\! + \!\frac{C_3}{m}\!\right]\nonumber \\  & + h\left\|[\nabla_{\x\x}f]^{-1}\nabla_{t\x} f - [\nabla_{\x\x}f^*]^{-1}\nabla_{t\x} f^*  \right\| .
\end{align}
We proceed to find an upper bound for the norm $\left\|[\nabla_{\x\x}f]^{-1}\nabla_{t\x} f - [\nabla_{\x\x}f^*]^{-1}\nabla_{t\x} f^*  \right\| $ in the right-hand side of \eqref{eq.imp00}. By adding and subtracting the term $[\nabla_{\x\x}f^*]^{-1}\nabla_{t\x} f$ and using triangle inequality we can write 
\begin{align}\label{eq.hessian_cross_term}
&\left\|[\nabla_{\x\x}f]^{-1}\nabla_{t\x} f - [\nabla_{\x\x}f^*]^{-1}\nabla_{t\x} f^*  \right\|  \nonumber\\
&\qquad  \leq \left\|[\nabla_{\x\x}f]^{-1}\nabla_{t\x} f - [\nabla_{\x\x}f^*]^{-1}\nabla_{t\x} f\right\|\nonumber\\ &\qquad\qquad+  \left\| [\nabla_{\x\x}f^*]^{-1}\nabla_{t\x} f- [\nabla_{\x\x}f^*]^{-1}\nabla_{t\x} f^*  \right\|.
\end{align}
We may bound the first and second-order derivative terms in \eqref{eq.hessian_cross_term} by using Assumption~\ref{as.smooth} regarding the functional smoothness as well as the strong convexity constant $m$ of the Hessian in Assumption \ref{as.str} to write
\begin{align}\label{eq.a}
&\left\|[\nabla_{\x\x}f]^{-1}\nabla_{t\x} f - [\nabla_{\x\x}f^*]^{-1}\nabla_{t\x} f^*  \right\|  \\
&\qquad  \leq C_0  \left\|[\nabla_{\x\x}f]^{-1} - [\nabla_{\x\x}f^*]^{-1}\right\| + \frac{1}{m}\left\| \nabla_{t\x} f- \nabla_{t\x} f^*  \right\|.\nonumber
\end{align}
We now further bound the first term of the right-hand side. To do that, we use the non-singularity of the Hessian to write 
\begin{multline}\label{eq.dummy10_midstep}
\left\|[\nabla_{\x\x}f]^{-1} - [\nabla_{\x\x}f^*]^{-1}\right\| = \\ \|[\nabla_{\x\x}f^*]^{-1}(\nabla_{\x\x}f - \nabla_{\x\x}f^*)[\nabla_{\x\x}f]^{-1} \|,
\end{multline} 
which by employing, once again, the strong convexity constant $m$ of the Hessian in Assumption \ref{as.str} we can bound as
\begin{equation}\label{eq.dummy10}
\left\|[\nabla_{\x\x}f]^{-1} - [\nabla_{\x\x}f^*]^{-1}\right\| \leq \frac{1}{m^2}\|\nabla_{\x\x}f - \nabla_{\x\x}f^* \|.
\end{equation} 
Substituting the upper bound in \eqref{eq.dummy10} for the norm $\left\|[\nabla_{\x\x}f]^{-1} - [\nabla_{\x\x}f^*]^{-1}\right\|$ into \eqref{eq.a} yields
\begin{align}\label{eq.amid}
&\left\|[\nabla_{\x\x}f]^{-1}\nabla_{t\x} f - [\nabla_{\x\x}f^*]^{-1}\nabla_{t\x} f^*  \right\| \nonumber \\
&\qquad   \leq \frac{C_0}{m^2}  \left\|\nabla_{\x\x}f - \nabla_{\x\x}f^*\right\| + \frac{1}{m}\left\| \nabla_{t\x} f- \nabla_{t\x} f^*  \right\|.
\end{align}
We consider the Taylor expansion of the second-order term in \eqref{eq.amid}, and apply the Mean Value Theorem with $\tilde{\x}$ as a point on the line between $\x_k$ and $\x^*(t_k)$ to obtain 
%
\begin{align} \label{eq.dummy11}
\left\| \nabla_{\x\x} f- \nabla_{\x\x} f^*  \right\| &
\leq \left\|\nabla_{\x \x\x} f(\tilde{\x}; t_k)\right\| \|\x_k - \x^*(t_k)\| \nonumber\\ 
&\leq C_1 \|\x_k - \x^*(t_k)\|.
\end{align}
Applying the same argument for the mixed second-order term implies 
\begin{align}
\left\| \nabla_{t\x} f- \nabla_{t\x} f^*  \right\| &
\leq \left\|\nabla_{\x t\x} f(\tilde{\x}; t_k)\right\| \|\x_k - \x^*(t_k)\| \nonumber \\
& \leq C_2 \|\x_k - \x^*(t_k)\| \label{eq.dummy111}
\end{align}
%
The expressions in \eqref{eq.dummy11} and \eqref{eq.dummy111} may be substituted together into~\eqref{eq.amid} to yield
\begin{align}\label{something_new}
&\left\|[\nabla_{\x\x}f]^{-1}\nabla_{t\x} f - [\nabla_{\x\x}f^*]^{-1}\nabla_{t\x} f^*  \right\|  \\
&\qquad\qquad\qquad \leq \left(\frac{C_0 C_1 }{m^2}+ \frac{C_2 }{m}\right) \|\x_k - \x^*(t_k)\|.
\nonumber
\end{align}
By substituting the upper bound in \eqref{something_new} into~\eqref{eq.imp00} and considering the definition of $\sigma$ in \eqref{thm:rho_sigma}, we obtain that
\begin{multline}\label{eq.fin1}
\|{\x}_{k+1|k} - \x^*(t_{k+1})\| \leq \sigma \|{\x}_{k} - \x^*(t_{k})\| + \\ \frac{h^2}{2}\!\left[\frac{C_0^2 C_1}{m^3} \!+ \!\frac{2 C_0 C_2}{m^2} \!+\! \frac{C_3}{m}\!\right] .
\end{multline}
For the correction step [cf. \eqref{GTT_correction}] , we may use the standard property of  gradient descent for strongly convex functions with Lipschitz gradients. In particular, the Euclidean error norm of the  gradient descent method converges as
\begin{equation}\label{eq.fin2}
\|{\hat{\x}}_{k+1}^{s+1} - \x^*(t_{k+1})\| \leq \rho \|{\hat{\x}}_{k+1}^{s} - \x^*(t_{k+1})\| .
\end{equation}
where $\rho = \max\{|1-\gamma m|,|1- \gamma L|\}$. To see this, it is sufficient to write the  gradient step as 
\begin{multline}\label{eq.explanation1}
\|{\hat{\x}}_{k+1}^{s+1} - \x^*(t_{k+1})\|  \\ =  \|{\hat{\x}}_{k+1}^{s} -\gamma \nabla_{\x} f(\hat{\x}_{k+1}^s; t_{k+1}) - \x^*(t_{k+1})\| .
\end{multline}
%
According to the optimality condition we can write $\nabla_{\x} f(\x^*(t_{k+1}); t_{k+1})=\mathbf{0}$. Considering this observation and the equality in \eqref{eq.explanation1} we obtain
\begin{align}\label{eq.explanation3}
\|{\hat{\x}}_{k+1}^{s+1}\! -\! \x^*(t_{k+1})\|&\!=\!   \|{\hat{\x}}_{k+1}^{s} - \x^*(t_{k+1}) \\
&   \hskip-15mm- \gamma[ \nabla_{\x} f(\hat{\x}_{k+1}^s; t_{k+1})-\nabla_{\x} f(\x^*(t_{k+1}); t_{k+1})]\|.\nonumber
\end{align}
Consider now the continuous function $g:\reals^n\times\reals_{+} \to \reals$ defined as $g(\x; t) := \x -\gamma \nabla_{\x} f(\x; t)$. Given the boundedness of the Hessian and the strong convexity of $f(\x; t)$, the gradient of $g(\x; t)$ is bounded as~\cite[page~13]{Ryu2015}
\begin{equation}\label{boundg}
\|\nabla_{\x} g(\x; t)\| \leq \max\{|1 -\gamma m|,|1 - \gamma L|\} = \rho, 
\end{equation}
for all $\x \in \mathbb{R}^n$. The bound~\eqref{boundg} implies that $g(\x; t)$ is Lipschitz, therefore we can upper bound~\eqref{eq.explanation3} as
\begin{multline}\label{eq.explanation4}
\|{\hat{\x}}_{k+1}^{s+1}\! -\! \x^*(t_{k+1})\| \!=\! \|g({\hat{\x}}_{k+1}^{s}; t_{k+1}) - g(\x^*(t_{k+1}); t_{k+1})\| \\ \!\leq\! \rho \|{\hat{\x}}_{k+1}^{s} \!-\! \x^*(t_{k+1})\|.
\end{multline}
Notice that the relation~\eqref{eq.explanation4} is equivalent to the claim in~\eqref{eq.fin2}. 

Observe that the sequence ${\hat{\x}}_{k+1}^{s}$ is initialized by the predicted variable $\x_{k+1|k}$ and the corrected variable $\x_{k+1}$ is equal to $ {\hat{\x}}_{k+1}^{\tau}$. Considering these observations and the relation in \eqref{eq.fin2} between two consecutive iterates of the sequence ${\hat{\x}}_{k+1}^{s}$ we can write 
\begin{equation}\label{eq.fin222}
\|{{\x}}_{k+1}- \x^*(t_{k+1})\| \leq \rho^\tau \|\x_{k+1|k} - \x^*(t_{k+1})\| .
\end{equation}
We are ready to consider the combined error bound achieved by the prediction-correction scheme. By plugging the correction error of \eqref{eq.fin222} into the prediction error of \eqref{eq.fin1} we obtain
\begin{equation}\label{eq.fin_t_k}
\|{\x}_{k+1} - \x^*(t_{k+1})\| \leq \rho^\tau \sigma \|{\x}_{k} - \x^*(t_{k})\| +  \rho^\tau \Gamma,
\end{equation}
where $\Gamma:=(h^2/2)[C_0^2 C_1/m^3 + 2 C_0 C_2/m^2 + C_3/m]$ is defined to simplify the notation. Notice that the relation between $\|{\x}_{k+1} - \x^*(t_{k+1})\|$ and $\|{\x}_{k} - \x^*(t_{k})\| $ in \eqref{eq.fin_t_k}  also holds true for $\|{\x}_{k} - \x^*(t_{k})\| $ and $\|{\x}_{k-1} - \x^*(t_{k-1})\| $, i.e., 
\begin{equation}\label{eq.fin_t_k_1}
\|{\x}_{k} - \x^*(t_{k})\| \leq \rho^\tau \sigma \|{\x}_{k-1} - \x^*(t_{k-1})\| +  \rho^\tau \Gamma.
\end{equation}
Substituting the upper bound in \eqref{eq.fin_t_k_1} for $\|{\x}_{k} - \x^*(t_{k})\|$ into \eqref{eq.fin_t_k} implies an upper bound for $\|{\x}_{k+1} - \x^*(t_{k+1})\| $ in terms of the norm difference for time $k-1$ as
\begin{equation}\label{eq.fin_t_k_new}
\|{\x}_{k+1} - \x^*(t_{k+1})\| \leq (\rho^\tau \sigma)^2 \|{\x}_{k-1} - \x^*(t_{k-1})\|
 +\rho^\tau \Gamma(\rho^\tau \sigma+1) .
\end{equation}
Now recursively apply the relationship \eqref{eq.fin_t_k} backwards in time to the initial time sample and use the same argument form \eqref{eq.fin_t_k} to \eqref{eq.fin_t_k_new} to write 
\begin{equation}\label{eq.fin_t_00}
\|{\x}_{k+1} - \x^*(t_{k+1})\| \leq (\rho^\tau \sigma)^{k+1} \|{\x}_{0} - \x^*(t_{0})\| + \rho^\tau \Gamma \sum_{i=0}^k (\rho^\tau\sigma)^i.
\end{equation}
Substituting $k+1$ by $k$ and simplifying the sum in \eqref{eq.fin_t_00} (remembering that $\rho^\tau \sigma < 1$) leads to
\begin{equation}\label{eq.fin_t_0000}
\|{\x}_{k}\! - \!\x^*(t_{k})\| \!\leq\! (\rho^\tau \sigma)^{k} \|{\x}_{0} - \x^*(t_{0})\| \!+ \!\rho^\tau \Gamma \!\left[\! \frac{1-(\rho^\tau\sigma)^k}{1-\rho^\tau\sigma}\!\right].
\end{equation}
Considering the result in \eqref{eq.fin_t_0000} and the definition for the constant $\Gamma$
, the result in \eqref{result1} follows.

To establish the result stated in \eqref{result2}, observe that in the worst case, we may upper bound the term $\|[\nabla_{\x\x}f]^{-1}\nabla_{t\x} f - [\nabla_{\x\x}f^*]^{-1}\nabla_{t\x} f^* \|$ in \eqref{eq.imp} by using the bounds in Assumption \ref{as.smooth} to obtain the right-hand side of the following expression 
\begin{align}\label{eq.worst_case}
\left\|[\nabla_{\x\x}f]^{-1}\nabla_{t\x} f - [\nabla_{\x\x}f^*]^{-1}\nabla_{t\x} f^*  \right\| 
 & \leq \frac{2 C_0}{m}.
\end{align}
Substituting the bound in~\eqref{eq.worst_case} into \eqref{eq.imp00} yields
\begin{align}\label{eq.a_subst}
\|{\x}_{k+1|k} - \x^*(t_{k+1})\| &\leq \|{\x}_{k} - \x^*(t_{k})\| + h\, \frac{2 C_0 }{m} \nonumber\\
&\quad
+\frac{h^2}{2}\left[\frac{C_0^2 C_1}{m^3} + \frac{2 C_0 C_2}{m^2} + \frac{C_3}{m}\right] .
\end{align}
To simplify the notation we define a new constant $\Gamma_2:= 2 h { C_0 }/{m} $ and we use again the definition $\Gamma:=(h^2/2)[C_0^2 C_1/m^3 + 2 C_0 C_2/m^2 + C_3/m]$. Considering this definition and observing the relation in \eqref{eq.fin222} we can write 
\begin{equation}\label{eq.sampling_error2}
\|{\x}_{k+1} - \x^*(t_{k+1})\| \leq \rho^\tau \|{\x}_{k} - \x^*(t_{k})\| + \rho^\tau (\Gamma_2 + \Gamma).
\end{equation}
Now recursively apply the relationship \eqref{eq.sampling_error2} backwards in time to the initial time sample and use the same argument from \eqref{eq.fin_t_k} to \eqref{eq.fin_t_0000} to write 
\begin{multline}\label{eq.rho_error_dependence}
\|{\x}_{k+1} - \x^*(t_{k+1})\| \leq \rho^{\tau(k+1)} \|{\x}_{0} - \x^*(t_{0})\| \\ +  \rho^\tau (\Gamma_2 + \Gamma) \left[ \frac{1-\rho^{\tau (k+1)}}{1-\rho^\tau}\right].
\end{multline}
Note that relation \eqref{eq.rho_error_dependence} shows an upper bound for $\|{\x}_{k+1} - \x^*(t_{k+1})\|$ in terms of the initial error $\|{\x}_{0} - \x^*(t_{0})\|$ and an extra error term for the bound of convergence. If we substitute $k+1$ by $k$ in \eqref{eq.rho_error_dependence} and recall the definition of $\Gamma_2$ and $\Gamma$
, then the result in \eqref{result2} follows.

For completeness, we show that $\rho < 1$ requires the stepsize to be selected as $\gamma < 2/L$, which therefore enforce a finite right-hand side in~\eqref{eq.rho_error_dependence}. Starting by the definition of $\rho$, we require 
\begin{equation}
\rho := \max\{|1-\gamma m|,|1-\gamma L|\} < 1.
\end{equation}
Solving this equation for $\gamma$ and recalling that $m \leq L$ by Assumptions~\ref{as.str} and \ref{as.smooth}, the condition $\gamma < 2/L$ follows. 
\qed

\section{Proof of Theorem~\ref{th.NTT_convg} }\label{ap.newton}

We consider once again the proof of Theorem~\ref{th.GTT_convg}, in particular Eq.~\eqref{eq.fin1} for $k = 0$, due to the prediction step. For the correction step, if we applied one time the Newton method, we would have
\begin{equation}\label{eq.dummy300}
\|\x_{1} - \x^*(t_{1})\| \leq  \frac{C_1}{2 m} \|\x_{1|0} - \x^*(t_{1})\|^2.
\end{equation}
We proceed to check the validity of~\eqref{eq.dummy300}. To do so, we first simplify the notations as 
\begin{align}\label{eq:thm1_defs2_dummy}
&\nabla_{\x\x}f_1 = \nabla_{\x\x}f(\x_{1|0}; t_1), &&\nabla_{\x} f_1 = \nabla_{\x} f(\x_{1|0}; t_1),\nonumber \\ 
&\nabla_{\x\x}f^*_1 = \nabla_{\x\x}f({\x}^*(t_{1}); t_1),  && \nabla_{\x} f^*_1 = \nabla_{\x} f({\x}^*(t_{1}); t_{1}). 
\end{align}
Considering the update of the  Newton method which is used in the correction step of NTT we can write
\begin{align}
\|\x_{1} - \x^*(t_{1})\| &=  
 \|\x_{1|0} -  \nabla_{\x\x}f_1^{-1}\nabla_{\x}f_1- \x^*(t_{1})\|, \label{eq.aryan1}.
\end{align}
%
By factoring the Hessian inverse $\nabla_{\x\x}f_1^{-1}$ and using the fact that the norm of a product is smaller than the product of the norms, we can show that the right-hand side of~\eqref{eq.aryan1} is bounded above as
\begin{align}
& \|\x_{1|0} -  \nabla_{\x\x}f_1^{-1}\nabla_{\x}f_1- \x^*(t_{1})\|
\nonumber\\
&
\qquad \qquad \leq \|\nabla_{\x\x}f_1^{-1}\|\|\nabla_{\x\x}f_1(\x_{1|0} - \x^*(t_{1})) - \nabla_{\x}f_1\| \label{eq.aryan3}.
\end{align}
Notice that the norm $\|\nabla_{\x\x}f_1^{-1}\|$ is bounded above by $1/m$ according to the strong convexity assumption. Further, the 
optimality conditions imply $\nabla_{\x} f^*_1 =\mathbf{0}$. These observations imply that we can rewrite \eqref{eq.aryan3} as
\begin{align}
& \|\x_{1|0} -  \nabla_{\x\x}f_1^{-1}\nabla_{\x}f_1- \x^*(t_{1})\|
\nonumber\\
&
\qquad  \leq \frac{1}{m}\|\nabla_{\x\x}f_1(\x_{1|0} - \x^*(t_{1})) - \left(\nabla_{\x}f_1- \nabla_{\x} f^*_1 \right)\| \label{eq.aryan300}.
\end{align}
Define $\r_1 = \x_{1|0} - \x^*(t_{1})$ and $\mathbold{\xi}(\tau)\! =\! \x^*(t_1) + \tau (\x_{1|0} - \x^*(t_1))$. We now use the fundamental theorem of calculus and the Lipschitz continuity of the Hessian (Assumption~\ref{as.smooth}) to upper bound the rightmost term of~\eqref{eq.aryan300} as
\begin{align}
\|\nabla_{\x\x}f_1 \r_1&-\left(\nabla_{\x}f_1- \nabla_{\x} f^*_1 \right)\| \nonumber\\
&\qquad = \Big\|\nabla_{\x\x}f_1 \r_1 - \int_{0}^1 \!\nabla_{\x\x}f(\mathbold{\xi}(\tau); t_1)\r_1\textrm{d} \tau\Big\| \nonumber \\ 
& \qquad  = \Big\|   \r_1 \int_{0}^1 \! \nabla_{\x\x}f_1 -  \nabla_{\x\x}f(\mathbold{\xi}(\tau); t_1)\textrm{d} \tau \Big\| \nonumber \\ 
&\qquad  \leq \| \r_1\| \int_{0}^1 \|\nabla_{\x\x}f_1-\nabla_{\x\x}f(\mathbold{\xi}(\tau); t_1)\|\textrm{d} \tau \nonumber \\ 
&\qquad  \leq C_1 \| \r_1\|^2 \!\int_{0}^1 \!(1-\tau) \textrm{d}\tau = \frac{C_1}{2} \| \r_1\|^2 \label{eq.aryan4}.
\end{align}
Notice that the first inequality in \eqref{eq.aryan4} is implied by the Cauchy-Schwarz inequality and the second inequality is true because of the Lipschitz continuity of the gradients with constant $C_1$. By plugging the bound~\eqref{eq.aryan4} into~\eqref{eq.aryan300} and recalling the definition $\r_1 = \x_{1|0} - \x^*(t_{1})$ we obtain that
\begin{align}
\left\|\x_{1|0} -  \nabla_{\x\x}f_1^{-1}\nabla_{\x}f_1- \x^*(t_{1})\right\|
 \leq \frac{C_1}{2m} \|  \x_{1|0} - \x^*(t_{1})\|^2 \label{eq.aryan3000}.
\end{align}
Combining the inequalities in \eqref{eq.aryan1} and \eqref{eq.aryan3000} follows the claim in \eqref{eq.dummy300}.

Now consider the case that $\tau $ steps of the  Newton method are applied in the correction step of the NTT algorithm. Then, the the error $\|\x_{1} - \x^*(t_{1})\|$ at step $t_1$ is bounded above as
\begin{equation}\label{eq.dummy30}
\|\x_{1} - \x^*(t_{1})\| \leq  \left(\frac{C_1}{2 m}\right)^{2\tau - 1}\|\x_{1|0} - \x^*(t_{1})\|^{2 \tau}.
\end{equation}
Notice that the upper bound for the prediction error in \eqref{eq.fin1} implies that the norm $\|{\x}_{1|0} - \x^*(t_{1})\|$ is bounded above as
\begin{equation}\label{some_bound_200}
\|{\x}_{1|0}\! - \!\x^*(t_{1})\|  \!\leq\!  \sigma \|{\x}_{0}\! - \!\x^*(t_{0})\|\!+\! \frac{h^2}{2}\!\left[\!\frac{C_0^2 C_1}{m^3} \!+\! \frac{2 C_0 C_2}{m^2} \!+ \!\frac{C_3}{m}\!\right].
\end{equation}
where $\sigma:= 1 + h\delta_1$, and $\delta_1$ is defined in \eqref{eq.simply}. Combining the inequalities in \eqref{eq.dummy30} and \eqref{some_bound_200} and considering the definitions $Q:=2m/C_1$ and $\delta_2:=C_0^2C_1/2 m^3 + {C_0 C_2}/{m^2} +{C_3}/{2m}$ yield 
\begin{equation}\label{eq.dummy3111}
\|\x_{1} - \x^*(t_{1})\|  \leq Q^{-(2\tau -1)}  \left(\sigma \|{\x}_{0} - \x^*(t_{0})\| + h^2 \delta_2\right)^{2\tau} .
\end{equation}
Based on the assumption in \eqref{locality} the initial error is bounded above by $ch^2$ (with $c$ an arbitrary positive constant). Substituting this upper bound into the right-hand side of \eqref{eq.dummy3111} follows
\begin{equation}\label{eq.dummy31}
\|\x_{1} - \x^*(t_{1})\|   \leq Q^{-(2\tau -1)} \left((\sigma c + \delta_2) h^2 \right)^{2\tau}.
\end{equation}
Notice that the inequality in \eqref{eq.dummy31} shows that the error $\|\x_{t} - \x^*(t_{t})\| $ for the step $t=1$ is in the order of $O(h^{4\tau})$ which is a better error bound with respect to the initial error $\|\x_{0} - \x^*(t_{0})\| =O(h^2)$. We now proceed to find under which conditions the error in inequality~\eqref{eq.dummy31} is valid for all $\|\x_{k} - \x^*(t_{k})\|$ with $k\geq 1$. To do so, we use induction. We first establish the sufficient conditions for which $\|\x_{1} - \x^*(t_{1})\|\leq c h^2$; then we substitute $\|\x_{2} - \x^*(t_{2})\|$ with $\|\x_{1} - \x^*(t_{1})\|$ and $\|\x_{1} - \x^*(t_{1})\|$ with $\|\x_{0} - \x^*(t_{0})\|$ in~\eqref{eq.dummy3111} and by induction on the error term $\|\x_{k} - \x^*(t_{k})\|$ we will prove the claim that $\|\x_{k} - \x^*(t_{k})\|=O(h^4)$ with $k\geq 1$. In particular, we need to make sure that the sampling period $h$ is chosen such that the upper bound in \eqref{eq.dummy31} is smaller than $ch^2$, i.e.,
\begin{equation}\label{eq.dummy32}
Q^{-(2\tau -1)} \left(\sigma c h^2 + \delta_2 h^2 \right)^{2\tau} \leq c h^2.
\end{equation}
Observe that according to the required condition for the sampling period $h$ in \eqref{up2} we can write $h\leq1$. Therefore, the constant $\sigma:= 1+ h\delta_1$ is bounded above by $1+ \delta_1$. Substituting $1+ \delta_1$ for $\sigma $ in \eqref{eq.dummy32} implies a sufficient condition for \eqref{eq.dummy32} as
\begin{equation}\label{eq.dummy3002}
Q^{-(2\tau -1)} \left((1+ \delta_1) c h^2 + \delta_2 h^2 \right)^{2\tau} \leq c h^2.
\end{equation}
We emphasize that if the inequality in \eqref{eq.dummy3002} holds true then the statement in \eqref{eq.dummy32} is satisfied. Regrouping the terms in \eqref{eq.dummy3002} leads to the following condition for the sampling interval $h$ as
\begin{align}\label{cond_on_h}
  h \leq \left[\frac{Q^{(2\tau -1)}c}{((1+ \delta_1) c \!+\! \delta_2)^{2\tau}}\right]^{\frac{1}{4\tau-2}}.
\end{align}
Therefore, if \eqref{cond_on_h} is satisfied then \eqref{eq.dummy3002} and subsequently \eqref{eq.dummy32} are satisfied. Based on the assumption in \eqref{up2}, we know that \eqref{cond_on_h} is valid and the condition in \eqref{eq.dummy32} is satisfied. This observation in conjunction with the inequality in \eqref{eq.dummy31} implies that 
\begin{equation}
\|\x_{1} - \x^*(t_{1})\| \leq ch^2.
\end{equation}
By starting again from~\eqref{eq.dummy3111}, and by substituting $\|\x_{2} - \x^*(t_{2})\|$ with $\|\x_{1} - \x^*(t_{1})\|$ and $\|\x_{1} - \x^*(t_{1})\|$ with $\|\x_{0} - \x^*(t_{0})\|$, we arrive at the inequality
\begin{equation}\label{eq.dummy31for2}
\|\x_{2} - \x^*(t_{2})\|   \leq Q^{-(2\tau -1)} \left((\sigma c + \delta_2) h^2 \right)^{2\tau}.
\end{equation}
Since the condition in~\eqref{cond_on_h} does not depend on the optimality gap, they yield $\|\x_{2} - \x^*(t_{2})\| \leq c h^2$. By applying the induction argument, we can now show that 
\begin{equation}\label{eq.dummy31fork}
\|\x_{k} - \x^*(t_{k})\|   \leq Q^{-(2\tau -1)} \left((\sigma c + \delta_2) h^2 \right)^{2\tau},
\end{equation}
for all $k\geq 1$, which is~\eqref{NewtonResult2}.
\qed

\section{Proof of Theorem~\ref{th.AGT_convg} }\label{ap.agt}


We prove Theorem~\ref{th.AGT_convg} by evaluating the extra error term coming from the approximate time derivative in~\eqref{fobd}. In particular, consider the Taylor's expansion of the gradient ${\nabla}_{\x}{f}(\x_{k}; t_{k-1})$ near the point $(\x_k,t_k)$ which is given by
\begin{multline}\label{taylor_series100}
{\nabla}_{\x}{f}(\x_{k}; t_{k-1}) =  {\nabla}_{\x}{f}(\x_k; t_k) - h\,{\nabla}_{t\x}{f}(\x_k; t_k) + \\  h^2/2\,  {\nabla}_{tt\x}{f}(\x_k; s) .
\end{multline} 
for a particular $s\!\in\![t_{k-1}, t_k]$. Regrouping the terms in \eqref{taylor_series100} it follows that the partial mixed gradient $ {\nabla}_{\x}{f}(\x_k; t_k) $ can be written as
\begin{multline}\label{taylor_series200}
{\nabla}_{t\x}{f}(\x_k; t_k) = \frac{{\nabla}_{\x}{f}(\x_k; t_k)\! -\! {\nabla}_{\x}{f}(\x_{k}; t_{k-1})}{h} 
+ \\ {h}/{2} \, {\nabla}_{tt\x}{f}(\x_k; s).
\end{multline}
Considering the definition of the approximate partial mixed gradient $\tilde{\nabla}_{t\x}f(\x_k; t_k) $ in~\eqref{fobd} and the expression for the exact mixed gradient $ {\nabla}_{\x}{f}(\x_k; t_k) $ in \eqref{taylor_series200}, we obtain that 
\begin{equation}\label{taylor_series300}
{\nabla}_{t\x}{f}(\x_{k}; t_{k}) -\tilde{\nabla}_{t\x}f(\x_k; t_k) =\frac{h}{2}  {\nabla}_{tt\x}{f}(\x_k; s).
\end{equation}
Based on Assumption \ref{as.smooth} the norm ${\nabla}_{tt\x}{f}(\x_k; s)$ is bounded above by $C_3$. Therefore, the error of the partial mixed gradient approximation is upper bounded by 
\begin{equation}\label{taylor_series400}
\|{\nabla}_{t\x}{f}(\x_{k}; t_{k}) -\tilde{\nabla}_{t\x}f(\x_k; t_k)\| \leq \frac{hC_3}{2}.
\end{equation}

Consider the approximate prediction step of the AGT algorithm in \eqref{appro_prediction}. By adding and subtracting the exact prediction direction $h [{\nabla}_{\x\x}{f}(\x_k; t_k)]^{-1} {\nabla}_{t\x}{f}_k $ to the right-hand side of the update in \eqref{appro_prediction} we obtain
\begin{align}\label{new_number}
\x_{k+1|k} &= \x_k - h \,[{\nabla}_{\x\x}{f}(\x_k; t_k)]^{-1} {\nabla}_{t\x}{f}(\x_k; t_k)  +\\ 
& 
\!\!\!\!\!\!\!\!+h\, [{\nabla}_{\x\x}{f}(\x_k; t_k)]^{-1} \!\! \left({\nabla}_{t\x}{f}(\x_k; t_k)  -\tilde{\nabla}_{t\x}f(\x_k; t_k)\right)\! . \nonumber 
\end{align}
Subtracting ${\x}^*(t_{k+1}) = {\x}^*(t_{k}) - h\, [\nabla_{\x\x}f^*]^{-1} \nabla_{t\x} f^* + \mathbold{\Delta}_k$ in \eqref{eq:null_residual} from \eqref{new_number}, and applying the triangle inequality lead to 
\begin{align}\label{eq.imp200}
&\|{\x}_{k+1|k} -  \x^*(t_{k+1})\| \leq \|{\x}_{k} - \x^*(t_{k})\|  \\
& \quad+ h\left\|[\nabla_{\x\x}f]^{-1}\nabla_{t\x} f - [\nabla_{\x\x}f^*]^{-1}\nabla_{t\x} f^*  \right\|  + \|\mathbold{\Delta}_k\| \nonumber \\
&\quad +h\left\| [{\nabla}_{\x\x}{f}(\x_k; t_k)]^{-1} \!\! \left({\nabla}_{t\x}{f}(\x_k; t_k)  -\tilde{\nabla}_{t\x}f(\x_k; t_k)\right)\right\| \nonumber.
\end{align}
Observe the upper bound for the norm $\|\mathbold{\Delta}_k\|$ in \eqref{prop_claim_err_bound}. Further, observe that $\left\| [{\nabla}_{\x\x}{f}(\x_k; t_k)]^{-1} \!\! \left({\nabla}_{t\x}{f}(\x_k; t_k)  -\tilde{\nabla}_{t\x}f(\x_k; t_k)\right)\right\|$ is bounded above by $C_3h/2m$ according to \eqref{taylor_series400} and Assumption \ref{as.smooth}. Substituting these upper bounds into \eqref{eq.imp200} yields
\begin{align}\label{eq.imp300}
\|{\x}_{k+1|k}& \!- \! \x^*(t_{k+1})\|\! \leq \!\|{\x}_{k}\! -\! \x^*(t_{k})\|\! \nonumber\\&\quad\quad\quad +\!\frac{h^2}{2}\!\left[\!\frac{C_0^2 C_1}{m^3} \!+\! \frac{2 C_0 C_2}{m^2} \!+ \!\frac{2 C_3}{m}\!\right] \nonumber\\  & + h\left\|[\nabla_{\x\x}f]^{-1}\nabla_{t\x} f - [\nabla_{\x\x}f^*]^{-1}\nabla_{t\x} f^*  \right\|  .
\end{align}
Observe that the inequality for the AGT algorithm in \eqref{eq.imp300} is identical to the result for the GTT method in \eqref{eq.imp00} except for the multiplier of $h^2$. This observation implies that by following the same steps from \eqref{eq.hessian_cross_term} to \eqref{eq.fin_t_0000} we can prove the claim in \eqref{result2a}. Likewise, if we redo the steps from \eqref{eq.worst_case} to \eqref{eq.rho_error_dependence}, the claim in \eqref{result1a} can be followed from the result in \eqref{eq.imp300}.
\qed

\section{Proof of Theorem~\ref{th.ANT_convg} }\label{ap.ant}


The proof of Theorem~\ref{th.ANT_convg} is based on the proof of Theorems~\ref{th.NTT_convg} and \ref{th.AGT_convg}. Since the correction step of NTT and ANT are identical, we can redo the steps from \eqref{eq.dummy300} to \eqref{eq.dummy30} to show that 
\begin{equation}\label{eq.dummy3000}
\|\x_{1} - \x^*(t_{1})\| \leq  Q^{-(2\tau - 1)}\|\x_{1|0} - \x^*(t_{1})\|^{2 \tau},
\end{equation}
where $Q=2m/C_1$. The prediction step of AGT and ANT are identical, therefore the result in~\eqref{eq.imp300} also holds true for ANT. Consider the result in ~\eqref{eq.imp300} for $k=0$. Using the inequality in \eqref{something_new} we can simplify the right-hand side of~\eqref{eq.imp300} as
\begin{equation}\label{eq.imp3120}
\|{\x}_{1|0} -  \x^*(t_{1})\| \!\leq\! \sigma\|{\x}_{0} - \x^*(t_{0})\|\!+\!\frac{h^2}{2}\!\left[\!\frac{C_0^2 C_1}{m^3} \!+\! \frac{2 C_0 C_2}{m^2}\!+\! \frac{2 C_3}{m}\!\right]\! ,
\end{equation}
where $\sigma=1+ h({C_0 C_1 }/{m^2}+ {C_2 }/{m})$. Combining the inequalities in \eqref{eq.dummy3000} and \eqref{eq.imp3120} and considering the definition of $\delta_2'$ in \eqref{zahra} lead to
\begin{equation}\label{eq.dummy312041}
\|\x_{1} - \x^*(t_{1})\|  \leq Q^{-(2\tau -1)}  \left(\sigma \|{\x}_{0} - \x^*(t_{0})\| + h^2 \delta_2'\right)^{2\tau} .
\end{equation}
The result for ANT in \eqref{eq.dummy312041} is similar to the result for NTT in \eqref{eq.dummy3111}. By following the steps from \eqref{eq.dummy31} to \eqref{eq.dummy31fork} the result in \eqref{NewtonResult222} follows.
\qed

\bibliographystyle{ieeetr}
\bibliography{PaperCollection2}

\end{document}